\documentclass[english,subeqn]{article}
\usepackage{helvet}
\usepackage[T1]{fontenc}
\usepackage[latin9]{inputenc}
\usepackage[a4paper]{geometry}
\usepackage{textcomp}
\usepackage{amsmath}
\usepackage{graphicx}
\usepackage{amssymb}
\usepackage{esint}

\makeatletter

\newcommand{\lyxmathsym}[1]{\ifmmode\begingroup\def\b@ld{bold}
  \text{\ifx\math@version\b@ld\bfseries\fi#1}\endgroup\else#1\fi}

\providecommand{\tabularnewline}{\\}

\newcommand{\lyxaddress}[1]{
\par {\raggedright #1
\vspace{1.4em}
\noindent\par}
}

\makeatother

\usepackage{babel}

\begin{document}

\title{Boundary finite size corrections for multiparticle states and planar
AdS/CFT}

\author{Zoltán Bajnok$^{a}$ and László Palla$^{b}$}

\maketitle

\lyxaddress{\begin{center}
\emph{$^{a}$Theoretical Physics Research Group of the Hungarian
Academy of Sciences,}\\
\emph{$^{b}$Institute for Theoretical Physics, Roland Eötvös University,
}\\
\emph{H-1117 P\'azm\'any s. 1/A, Budapest, Hungary}
\par\end{center}}

\vspace{-8cm}

\hspace{11cm}ITP-Budapest Report 647

\vspace{8cm}
\begin{abstract}
We propose formulas for the Lüscher type finite size energy correction
of multiparticle states on the interval and evaluate them for the
simplest case in the AdS/CFT setting. By this we determine the leading
wrapping correction to the anomalous dimension of the simplest determinant
type operator, which corresponds to a one particle state on the $Y=0$
brane. 
\end{abstract}

\section{Introduction}

In the last few years much progress has been made in computing the
spectrum of anomalous dimensions of planar $\mathcal{N}=4$ supersymmetric
$SU(N)$ Yang-Mills theory (SYM). The progress relied partly on the
AdS/CFT correspondence \cite{Maldacena:1997re,Gubser:1998bc,Witten:1998qj}
between this theory and the type IIB closed string theory on the $AdS_{5}\times S^{5}$
background and partly on the integrability properties of both theories
\cite{Minahan:2002ve,Beisert:2003tq,Bena:2003wd,Kazakov:2004qf,Arutyunov:2004vx,Staudacher:2004tk,Beisert:2005tm}.
In particular it is accepted that the spectrum of anomalous dimensions
of single trace operators containing an asymptotically large number
of elementary SYM fields (which corresponds to the energy spectrum
of strings moving freely on $AdS_{5}\times S^{5}$ with large angular
momentum) is fully determined by the system of asymptotic Bethe Ansatz
equations (ABA) \cite{Beisert:2005fw}.

In extending these computations to operators with finite length {}``wrapping
effects'' \cite{Beisert:2004hm}, missing from the ABA, must be taken
into account. On the string theory side they are related to vacuum
polarization effects and, as was shown by Lüscher \cite{Luscher:1986pf},
can be described by the infinite volume scattering data. The Lüscher
correction is just the leading term of a systematical expansion, which
is summed up in the Thermodynamic Bethe Ansatz (TBA) program. The
idea that the TBA approach can be applied to the superstring sigma
model was advocated in \cite{Ambjorn:2005wa}. The appropriate generalization
of Lüscher's idea managed to describe the four \cite{Bajnok:2008bm}
and five \cite{Bajnok:2009vm} loop corrections to the anomalous dimension
of the Konishi operator in complete agreement with the direct gauge
theory computations, whenever they were available \cite{Fiamberti:2007rj,Fiamberti:2008sh,Velizhanin:2008jd}.
This idea proved to be very useful to calculate the anomalous dimensions
of various operators in the perturbative regime \cite{Bajnok:2008qj,Beccaria:2009eq,Beccaria:2009hg,Beccaria:2009vt,Lukowski:2009ce,Velizhanin:2010cm}.
A valid description to all couplings is based on the TBA description
\cite{Gromov:2009tv,Arutyunov:2009zu,Bombardelli:2009ns,Arutyunov:2009ur,Gromov:2009bc,Arutyunov:2009ax,Cavaglia:2010nm},
which, as it is expected, in the weakly coupled regime reproduces
the results of the generalized Lüscher correction \cite{Arutyunov:2010gb,Balog:2010xa,Balog:2010vf,Arutyunov:2010gu}. 

Parallel to these developments the study of finite size effects for
determinant type operators or equivalently for open strings has also
been started. After exhibiting the classical \cite{Mann:2006rh} and
weak coupling \cite{Berenstein:2005vf} integrability of various open
string models Hofman and Maldacena argued \cite{Hofman:2007xp} that
open strings on the $AdS_{5}\times S^{5}$ background attached to
the $Y=0$ or to the $Z=0$ giant graviton branes are integrable at
all values of the coupling. The boundary version of the asymptotic
BA have been worked out in \cite{Galleas:2009ye,Nepomechie:2009zi}
for the $Y=0$ brane, while in \cite{Correa:2009dm} for the $Z=0$
brane. 

Using the AdS/CFT generalization \cite{Palla:2008zc} of the boundary
state formalism \cite{Ghoshal:1993tm} the authors of \cite{Correa:2009mz}
adopted the Lüscher type boundary finite size energy correction \cite{Bajnok:2004tq,Bajnok:2006dn}
for the worldsheet QFT on a strip with width $L$ to study the ground
states of the $Y=0$ and $Z=0$ branes. As the $Y=0$ ground-state
is left invariant by some supersymmetry transformation the finite
size energy corrections vanish for all order of the perturbation theory,
showing also that the corresponding determinant operator is protected.
In contrast, in the $Z=0$ setting the vacuum transforms nontrivially
under the symmetry leaving a room for finite size corrections. The
authors of \cite{Correa:2009mz} managed to calculate the leading
wrapping type boundary correction and also checked against their direct
gauge theory results. The aim of the present paper is to extend their
results for excited states. 

In general we aim to derive the multiparticle generalization of Lüscher's
formula for the boundary setting and apply them for magnons reflecting
on the $Y=0$ brane. In the scattering description the $Y=0$ brane
is much simpler than the $Z=0$ brane since the corresponding reflection
factor is diagonal, opposed to the other case, where it is highly
nondiagonal. In particular we would like to evaluate Lüscher's correction
to the energy of a single magnon moving freely on a strip of width
$L$ reflecting on the boundaries. The Lüscher correction to the energy
gives the leading exponential correction in $L$, and can naturally
be converted into the leading wrapping correction to the anomalous
dimension of the corresponding operator. 

The paper is organized as follows: In the next Section we introduce
the gauge invariant determinant-type operators, whose anomalous dimensions
we are aiming to calculate. We will focus on a single impurity type
operator which correspond to a one-particle state on the string side
reflecting between two boundaries. We calculate its anomalous dimension
in two different ways: from the spin-chain description originating
from perturbative Feynman diagrams and from the integrable asymptotical
boundary Bethe Ansatz. As the boundary BA is asymptotical we analyze
next its Lüscher type correction. We propose expressions for the leading
Lüscher-type energy corrections in Section 3 for both relativistic
and non-relativistic theories. The non-relativistic expressions are
then evaluated at leading order in Section 4 for the simplest nontrivial
operator. Finally we conclude in Section 5 and outline some open problems.
Some technical details on how we determined the boundary state is
put into Appendix A, while in Appendix B we recall the explicit S-matrix
elements we used to calculate the finite size corrections. Finally,
for sake of completeness, we give the full weak coupling solution
of the one particle boundary BY equation in Appendix C.

\section{Determinant-type operators and the BBY equation}

In this section we present the operator in the gauge theory description
whose anomalous dimensions we are going to calculate. We calculate
its perturbative anomalous dimension both from the dilatation operator
and from the boundary Bethe-Yang equation of the string theory description.

\subsection{Gauge theory description}

In the gauge theory description the ground state of the $Y=0$ brane
corresponds to the operator \[
\epsilon_{i_{1}\dots i_{N-1}i_{N}}^{j_{1}\dots j_{N-1}j_{N}}Y_{j_{1}}^{i_{1}}\dots Y_{j_{N-1}}^{i_{N-1}}(Z^{J})_{j_{N}}^{i_{N}}\]
 while the excitations we consider correspond to replacing one of
the $Z$-s by an impurity $\chi$: \begin{equation}
{\cal O}_{Y}(Z^{k}\chi Z^{J-k-1})=\epsilon_{i_{1}\dots i_{N-1}i_{N}}^{j_{1}\dots j_{N-1}j_{N}}Y_{j_{1}}^{i_{1}}\dots Y_{j_{N-1}}^{i_{N-1}}(Z^{k}\chi Z^{J-k-1})_{j_{N}}^{i_{N}}\label{impurity}\end{equation}
 The sets of fields inside $(\dots)_{j_{N}}^{i_{N}}$ constitute the
states of the open spin chain. We keep the length of the chain (the
number of fields inside $(\dots)_{j_{N}}^{i_{N}}$ ) finite.

The ground state for finite $N$ is not supersymmetric \cite{Balasubramanian:2002sa}.
In the planar limit ($N\to\infty$) however, what we are analyzing
in this paper, integrability shows up and the supersymmetry of the
ground-state seems to be restored. As a consequence, the anomalous
dimension of the corresponding operator vanishes: indeed this was
shown perturbatively up to two loops in \cite{Hofman:2007xp} while
at the level of the leading exponential corrections in \cite{Correa:2009mz}.

We consider the $SU(2)$ sector first. In this case the impurity is
given by $Y$. The fields at the two ends of the chain cannot be $Y$'s
since the operator then would factorize into a determinant and a single
trace. Therefore $Y$ can only occupy the 'internal' positions of
the chain; to describe this we denote the first position of the chain
by index $0$ and the last one by $J-1$ when the $Y$ can occupy
$J-2$ different positions. Of course the total length of the chain
is $J$. Furthermore we introduce the abbreviation ${\cal O}_{Y}(Y_{j})$
for an operator of the form in (\ref{impurity}) with $Y$ standing
at the $j$-th position in $(Z^{j-1}YZ^{J-j})_{j_{N}}^{i_{N}}$.

The final form of the integrable two loop Hamiltonian in the $SU(2)$
subsector of the $Y=0$ brane is given in \cite{Hofman:2007xp}: \[
H=(2g^{2}-8g^{4})\sum\limits _{i=1}^{J-3}(\mathrm{I}-P_{i,i+1})+2g^{4}\sum\limits _{i=1}^{J-4}(\mathrm{I}-P_{i,i+2})+(2g^{2}-4g^{4})(q_{1}^{Y}+q_{J-2}^{Y})+2g^{4}(q_{2}^{Y}+q_{J-3}^{Y})\]
 where $P_{i,k}$ is the permutation operator between sites $i$ and
$k$ and $q_{i}^{Y}$ acts as the identity if the field at the $i$-th
position is $Y$ and as zero if it is not.

The shortest conceivable string accommodating a one particle excitation
is $ZYZ$ having $J=3$. It is clear that only the third term of $H$
has a non trivial action on ${\cal O}_{Y}(Y_{1})$ and the corresponding
energy eigenvalue is \[
\Delta_{3}=4g^{2}-8g^{4}\]
As will see later this corresponds to a particle of momentum $p=\frac{\pi}{2}$
, see (\ref{eq:Delta_n}) below. 

For $J=4$ one can look for the eigenstates of the Hamiltonian in
the form $\psi(1){\cal O}_{Y}(Y_{1})+\psi(2){\cal O}_{Y}(Y_{2})$.
In this case also the first and last terms of $H$ give contributions
and the eigenvalue equations have the form: \[
H\Psi=\Delta_{4}\Psi=\left(\begin{array}{cc}
4g^{2}-10g^{4} & -2g^{2}+8g^{4}\\
-2g^{2}+8g^{4} & 4g^{2}-10g^{4}\end{array}\right)\left(\begin{array}{c}
\psi(1)\\
\psi(2)\end{array}\right)\]
Comparing the solution of the eigenvalue equation to \cite{Hofman:2007xp}
using $\psi(j)\sim\sin(jp)$ $j=1,2$ we see that one gets the first
eigenvalue when $\psi(1)=\psi(2)$ i.e. when $p=\pi/3$ and then \[
\Delta_{4}^{+}=2g^{2}-2g^{4}=8g^{2}\sin^{2}(\pi/6)-32g^{4}\sin^{4}(\pi/6)\]
In a similar way the condition to get the second eigenvalue is $\psi(1)=-\psi(2)$
yielding $p=2\pi/3$ and also in this case \[
\Delta_{4}^{-}=6g^{2}-18g^{4}=8g^{2}\sin^{2}(\pi/3)-32g^{4}\sin^{4}(\pi/3)\]

\subsection{Boundary Bethe Yang equations}

Having only one impurity in the chain of $Z$-s corresponds to an
excitation (magnon) moving freely between two boundaries and reflecting
on them. The anomalous dimension of an operator of the form (\ref{impurity})
is related to the bulk energy $E(p)=\sqrt{1+16g^{2}\sin^{2}(\frac{p}{2})}$
of the magnon as \begin{equation}
\Delta_{n}=E(p_{n})-1=8g^{2}\sin^{2}(\frac{p_{n}}{2})-32g^{4}\sin^{4}(\frac{p_{n}}{2})+\dots\label{eq:Delta_n}\end{equation}
where $p_{n}$ are the discrete values of the momenta restricted by
the BBY equation and the set of $\Delta_{n}$ should coincide with
the (expansion of the) energy eigenvalues of the spin chain Hamiltonian
with the two boundaries.

To describe the BBY equation we consider two boundaries labeled by
$\alpha,\beta$ at a distance $L$ and a particle (magnon) that propagates
freely between them while undergoing nontrivial reflections at the
two ends. The reflections of a magnon with momentum $p$, carrying
the fundamental representation of $su(2\vert2)\otimes su(2\vert2)$,
is described by the following matrices \[
\mathbb{R}_{\alpha}(-p)=\mathbb{R}_{\beta}(p)=\mathbb{R}(p)=R_{0}(p)\mbox{diag}\left(-e^{i\frac{p}{2}},e^{-i\frac{p}{2}},1,1\right)\otimes\mbox{diag}\left(-e^{i\frac{p}{2}},e^{-i\frac{p}{2}},1,1\right)\]
 where the first two components correspond to the bosonic while the
last two, (in which the reflection is trivial), to the fermionic components
of the representation. The scalar factor can be determined from the
boundary crossing unitarity equation \cite{Hofman:2007xp,Chen:2007ec,Ahn:2008df}
to be \begin{equation}
R_{0}(p)=-e^{-ip}\sigma(p,-p)\label{eq:R0p}\end{equation}
 where $\sigma$ stands for the dressing phase \cite{Arutyunov:2009kf},
and $e^{-ip}$ is a CDD factor which we fixed from the weak coupling
limit. 

The BBY equation encodes the periodicity of the particle's wavefunction%
\footnote{In the AdS conventions we use the inverse of the scattering matrix
(compared to the relativistic case) in writing the ABA equations $e^{ip_{j}L}=\prod_{k}\mathbb{S}(p_{j},p_{k})$.
To keep this convention we also write the boundary ABA into this convention. %
}: \begin{equation}
e^{-2ipL}\mathbb{R}_{\alpha}(-p)\mathbb{R}_{\beta}(p)\equiv e^{-2ip(L+1)}\sigma(p,-p)^{2}\mbox{diag}(e^{ip},e^{-ip},1,1)\otimes\mbox{diag}(e^{ip},e^{-ip},1,1)=1\label{eq:bby}\end{equation}
 Clearly the two pieces of the scalar factor effect the BBY equation
and consequently the allowed momenta of the reflecting magnons in
a different way: while up to $g^{6}$ one can forget about the dressing
factor, the exponential factor effectively shifts the width of the
strip by one unit. We analyze the general solution of the equation
in Appendix C.

Here we analyze the weak coupling solutions of the BBY (\ref{eq:bby})
up to the order of $g^{4}$. Since $p$ must be in the range $0<p<\pi$,
its allowed values for a magnon with labels $(11)$, which corresponds
to the $Y$ type impurity are as follows: \[
p_{n}=n\frac{\pi}{L},\quad n=1,\dots L-1\]

Now we can compare these momenta with the ones obtained from analysing
the spin chain Hamiltonian. We can see that if  $L=J-1$ then the
two sets of energies are identical. (We also verified this for the
$SU(3)$ subsector spanned by the three scalar fields $W$, $Z$ and
$Y$). This way we demonstrated that the weak coupling limit of the
solutions of the BBY equations matches with the results of the spin
chain calculations in both the $SU(2)$ and in the $SU(3)$ sectors.

\subsection{The dressing factor and higher order weak coupling solutions}

In the higher orders in $g$ the dressing factor also effects the
solutions of the BBY. Using the explicit form presented in \cite{Beisert:2006ez}
one finds \[
\sigma(p,-p)=e^{-ig^{6}2^{8}\zeta(3)\sin^{5}\left(\frac{p}{2}\right)\cos\left(\frac{p}{2}\right)+{\cal O}(g^{8})}=1-ig^{6}2^{8}\zeta(3)\sin^{5}\left(\frac{p}{2}\right)\cos\left(\frac{p}{2}\right)+{\cal O}(g^{8})\]
 Writing the momentum of the magnon with labels $(11)$ as \[
p=p_{n}+\delta p=n\frac{\pi}{L}+\delta p\]
 in (\ref{eq:bby}) yields \[
\delta p=-g^{6}\frac{2^{8}}{L}\zeta(3)\sin^{5}\left(\frac{p_{n}}{2}\right)\cos\left(\frac{p_{n}}{2}\right)+{\cal O}(g^{8})\]
 Since in the dispersion relation the momentum dependence is multiplied
by $g^{2}$ , the shift in $\delta p$ effects only the 8-th order
term: \[
E(p_{n})-1=8g^{2}\sin^{2}\left(\frac{p_{n}}{2}\right)-32g^{4}\sin^{4}\left(\frac{p_{n}}{2}\right)+256g^{6}\sin^{6}\left(\frac{p_{n}}{2}\right)\]
 \[
\qquad\quad\ \ -g^{8}\bigl(2560\sin^{8}\left(\frac{p_{n}}{2}\right)+\frac{2^{11}}{L}\zeta(3)\sin^{6}\left(\frac{p_{n}}{2}\right)\cos^{2}\left(\frac{p_{n}}{2}\right))+\dots\]
 Thus for a $(11)$ magnon on the shortest possible strip - i.e. when
$L=2$ ($p_{1}=\pi/2$) we find \[
E\left(\frac{\pi}{2}\right)-1=4g^{2}-8g^{4}+32g^{6}-g^{8}(160+64\zeta(3))+\dots\]

The BBY equations (\ref{eq:bby}) determine the \emph{polynomial}
finite size corrections only and next we consider the leading \emph{exponential}
corrections (known as Lüscher corrections). In particular we are interested
in the corrections of the energy of a one particle state. The Lüscher
corrections to the ground state energies for both the $Y=0$ and the
$Z=0$ branes were determined already in \cite{Correa:2009mz}).

The analytic expression for the Lüscher correction to the energy of
any excited state is not known in general even in the case of ordinary
relativistic theories. Therefore, first we propose such a general
expression which we extract/conjecture from studying certain excited
state TBA and NLIE equations derived in some specific relativistic
boundary theories. Then we generalize this to the AdS/CFT context
on the $Y=0$ brane.

\section{Boundary finite size corrections for multiparticle states}

In this Section we propose expressions for the leading finite size
corrections to the BBY energy of multiparticle states on the strip.
After reviewing the analog proposal for multiparticle states on the
circle we formulate our conjecture in the relativistic boundary setting.
 Having confirmed the proposed expressions against exactly known integral
equations we extend them to the non-relativistic realm valid also
for the AdS/CFT correspondence.

\subsection{Integrable systems on the circle}

Suppose we analyze a system of particles interacting via a relativistically
invariant integrable interaction.

\subsubsection*{Infinite volume characteristics}

In infinite volume the system is characterized by the dispersion relation\[
E(p_{i})=\sqrt{m_{i}^{2}+p_{i}^{2}}\qquad\leftrightarrow(E(\theta_{i}),p(\theta_{i}))=m_{i}(\cosh\theta_{i},\sinh\theta_{i})\]
 and the factorized scattering matrix 

\begin{center}
$\mathbb{S}(\theta_{1}-\theta_{2})=S_{ij}^{kl}(\theta_{1}-\theta_{2})$~~~~~~~~~\includegraphics[height=1.5cm]{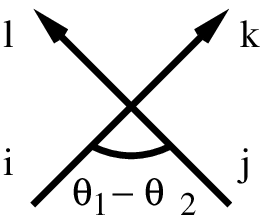}
\par\end{center}

\noindent which satisfies unitarity

\begin{center}
\includegraphics[height=3cm]{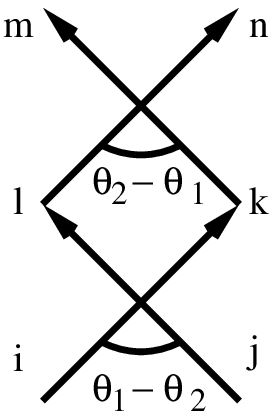}~~~~~~\includegraphics[height=3cm]{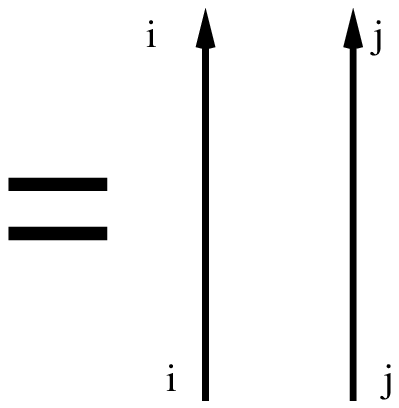}\[
S_{ij}^{kl}(\theta_{1}-\theta_{2})S_{lk}^{nm}(\theta_{2}-\theta_{1})=\delta_{i}^{m}\delta_{j}^{n}\]

\par\end{center}

\noindent crossing symmetry

\begin{center}
\includegraphics[height=1.5cm]{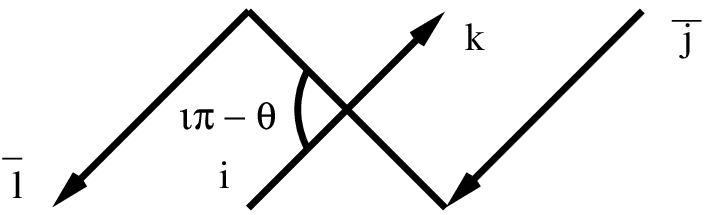}~~~~\includegraphics[height=1.5cm]{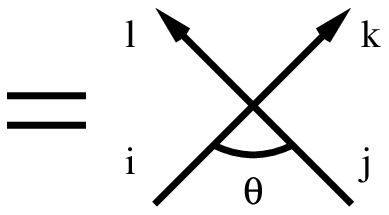}
\par\end{center}

\[
S_{\bar{l}i}^{\bar{j}k}(i\pi-\theta)=S_{ij}^{kl}(\theta)\]
and the Yang-Baxter equation

\begin{center}
\includegraphics[height=2cm]{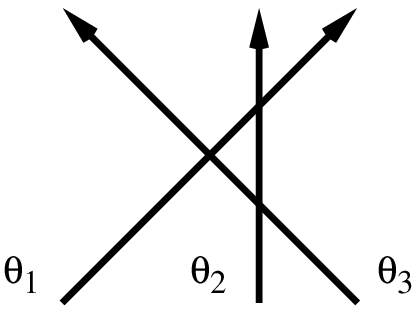}~~\includegraphics[height=2cm]{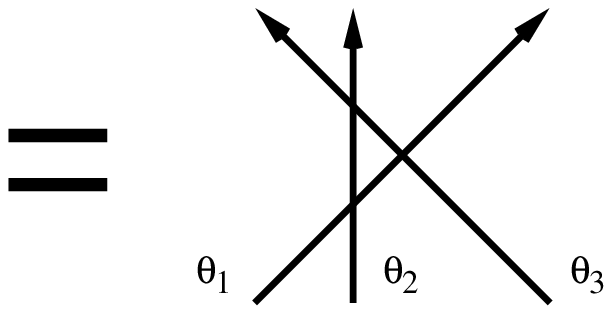}
\par\end{center}

\[
\mathbb{S}(\theta_{2}-\theta_{3})\mathbb{S}(\theta_{1}-\theta_{3})\mathbb{S}(\theta_{1}-\theta_{2})=\mathbb{S}(\theta_{1}-\theta_{2})\mathbb{S}(\theta_{1}-\theta_{3})\mathbb{S}(\theta_{2}-\theta_{3})\]

\subsubsection*{Asymptotically large volume spectrum}

In finite but large volume, $L$, the energy level of an $N$ particle
state can be described up to exponentially small corrections as the
sum of the individual energies \[
E(\theta_{1},\dots,\theta_{N})=\sum_{i}m_{i}\cosh\theta_{i}\]
Here the rapidities are constrained by the Bethe-Yang (BY) equations

\begin{center}
\includegraphics[height=2cm]{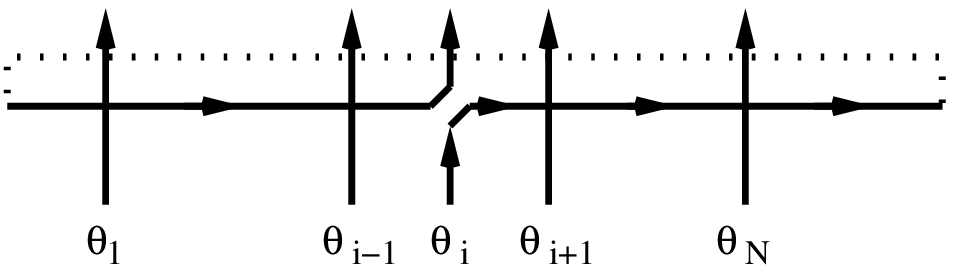}
\par\end{center}

\[
e^{ip(\theta_{i})L}\mathbb{S}(\theta_{i}-\theta_{i+1})\dots\mathbb{S}(\theta_{i}-\theta_{N})\mathbb{S}(\theta_{i}-\theta_{1})\dots\mathbb{S}(\theta_{i}-\theta_{i-1})=\mathbb{I}\quad;\qquad i=1,\dots,N\]
where $\mathbb{S}$ denotes the full $S$-matrix and their product
has to be diagonalized. If the $S$-matrix is regular, i.e. reduces
to the permutation operator at vanishing arguments

\noindent \begin{center}
$\mathbb{S}(0)=-\mathbb{P}$,~~~$S_{ij}^{kl}(0)=-\delta_{i}^{l}\delta_{j}^{k}$
~~~~~~~~~~~~~~~\includegraphics[height=1.5cm]{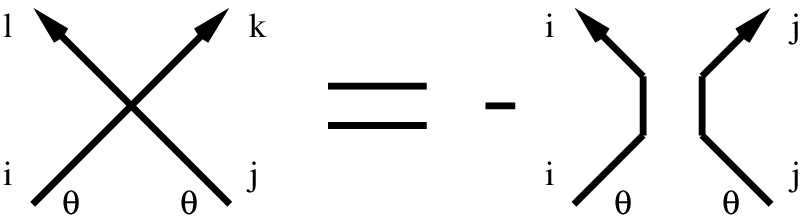}
\par\end{center}

\noindent then the BY equation can be nicely formulated in terms of
the asymptotic $Y$ function which is defined by means of the transfer
matrix

\begin{center}
\includegraphics[height=2cm]{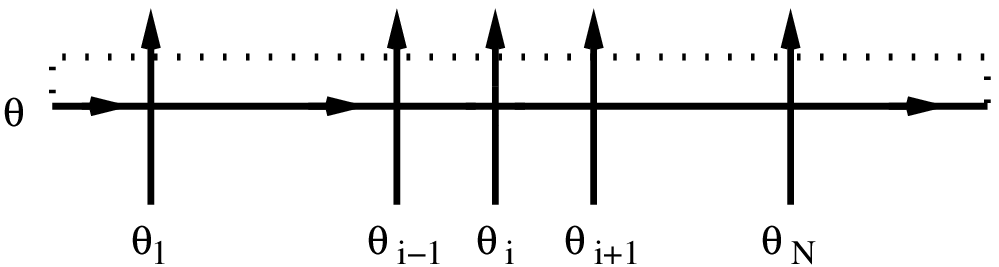}
\par\end{center}

\[
\mathbb{T}(\theta\vert\theta_{1},\dots,\theta_{N})=\mbox{Tr}(\mathbb{S}(\theta-\theta_{1})\dots\mathbb{S}(\theta-\theta_{N}))\]
If we denote the eigenvalue of the transfer matrix by $t(\theta\vert\theta_{1},\dots,\theta_{N})$
then the corresponding $Y$ function can be written as \[
Y_{as}(\theta\vert\theta_{1},\dots,\theta_{N})=e^{ip(\theta)L}t(\theta\vert\theta_{1},\dots,\theta_{N})\]
and the BY equation takes a particularly simple form \[
Y_{as}(\theta_{i}\vert\theta_{1},\dots,\theta_{N})=-1\]
The asymptotic $Y$-function is also relevant in describing the leading
exponentially small correction to the BY energies as we recall now.

\subsubsection*{Lüscher-type finite size corrections}

The BY energies contain all polynomial finite size corrections in
the inverse of the volume \cite{Luscher:1986pf} and there is a systematic
expansion for the additional exponentially small corrections which
are organized according to their exponents. The leading corrections
contain two terms: the so called integral or $F$-term and the residue
or $\mu$-term \cite{Luscher:1985dn}. As the $\mu$ term is simply
the residue of the integral term in the following we focus on the
integral term only. In \cite{Bajnok:2008bm} a formula was proposed
to describe the integral term of the leading exponential correction.
It consists of two parts, the first directly changes the energy in
the form\[
\Delta E=-\int_{-\infty}^{\infty}\frac{d\theta}{2\pi}\partial_{\theta}p(\theta)\, Y_{as}(\theta+i\frac{\pi}{2}\vert\theta_{1},\dots,\theta_{N})\]
and contains the vacuum polarization effects. The other one describes
how the finite volume vacuum changes the momentum quantization (or
other words the BY) equations: \[
\log Y_{as}(\theta_{i}\vert\theta_{1},\dots,\theta_{N})-\pi(2n+1)=-\partial_{\theta_{i}}\int_{-\infty}^{\infty}\frac{d\theta}{2\pi}\, Y_{as}(\theta+i\frac{\pi}{2}\vert\theta_{1},\dots,\theta_{N})\]
These formulas have been tested against the available exact integral
equations in diagonal theories like sinh-Gordon and Lee-Yang models
in \cite{Bajnok:2008bm} and for non-diagonal theories in \cite{Gromov:2008gj,Balog:2009ze,Balog:2010vf}
and in both cases perfect agreement have been found.

\subsubsection*{Non-relativistic models}

If the system is not relativistically invariant in infinite volume
then the above formulas have to be modified. The main difference is
that the scattering matrix no longer depends on the difference of
the (generalized) rapidities $u_{i}$, rather, it depends individually
on the two rapidities $\mathbb{S}(u_{i},u_{j})$. Still unitarity
$\mathbb{S}(u_{1},u_{2})=\mathbb{S}(u_{2},u_{1})^{-1}$, crossing
symmetry $\mathbb{S}^{c_{1}}(u_{1},u_{2})=\mathbb{S}(u_{2},u_{1}-\omega)$
(for some crossing parameter $\omega$) and regularity ($\mathbb{S}(u,u)=-\mathbb{P}$)
are supposed. Here $\mathbb{S}^{c_{1}}$ is charge conjugated in the
first particle only and in the relativistic case $\omega=i\pi$.

In describing the finite volume spectrum similar formulas can be introduced
as in the relativistic case. The transfer matrix is defined to be
\[
\mathbb{T}(u\vert u_{1},\dots,u_{N})=\mbox{Tr}(\mathbb{S}(u,u_{1})\dots\mathbb{S}(u,u_{N}))\]
and with its eigenvalue $t(\theta\vert\theta_{1},\dots,\theta_{N})$
the corresponding asymptotic $Y$ function can be written as \[
Y_{as}(u\vert u_{1},\dots,u_{N})=e^{ip(u)L}t(u\vert u_{1},\dots,u_{N})\]
From the regularity of the scattering matrix the BY equation follows
\[
Y_{as}(u_{i}\vert u_{1},\dots,u_{N})=-1\]
Moreover, the direct finite size energy correction has a similar form
as we have in the relativistic case: \[
\Delta E=-\int_{-\infty}^{\infty}\frac{du}{2\pi}\partial_{u}\tilde{p}(u)\, Y_{as}(u+\frac{\omega}{2}\vert u_{1},\dots,u_{N})\]
where the rapidity variable $u$ has been analytically continued into
its mirror domain: $u\to u+\frac{\omega}{2}$. The mirror theory can
be obtained from the original theory as follows: first we define the
Euclidean version of the model by analytically continuing in the time
variable $t=iy$ and considering space $x$ and imaginary time $y$
on an equal footing. This Euclidean theory can be considered as an
analytical continuation of another theory, in which $x$ serves as
the analytically continued time $x=i\tau$ and $y$ is the space coordinate.
The theory defined in terms of $y,\tau$ is called the mirror theory
and its dispersion relation can be obtained by the same analytical
continuation $E=i\tilde{p}$ and $p=i\tilde{E}$ see \cite{Arutyunov:2007tc}
in the AdS/CFT setting. In the general rapidity formulation we suppose
the mirror theory can be described by the $u\to u+\frac{\omega}{2}$
shift: $\tilde{E}(u)=-ip(u+\frac{\omega}{2})$ and $\tilde{p}(u)=-iE(u+\frac{\omega}{2})$. 

What is really difficult to figure out is the modification of the
BA. In the paper \cite{Bajnok:2008bm} a special case was analyzed
and the following form was proposed \[
\log Y_{as}(u_{i}\vert u_{1},\dots,u_{N})-\pi(2n+1)=\int_{-\infty}^{\infty}\frac{du}{2\pi}\, t^{'}(u+\frac{\omega}{2}\vert u_{1},\dots,u_{N})e^{ip(u+\frac{\omega}{2})L}\]
where $t^{'}(u\vert u_{1},\dots,u_{N})$ is the eigenvalue of $\mbox{Tr}(\mathbb{S}(u,u_{1})\dots(\partial_{u}\mathbb{S}(u,u_{i}))\dots\mathbb{S}(u,u_{N}))$
which was supposed to act diagonally on the multiparticle state whose
energy correction we are calculating. These formulas%
\footnote{To take into account the supersymmetric nature of the model the trace
$\mbox{Tr()}$has to be replaced by the supertrace $\mbox{sTr()}$.%
} have been used in the AdS/CFT realm at five loop \cite{Bajnok:2009vm,Lukowski:2009ce}
and compared to the TBA equations in \cite{Balog:2010xa,Balog:2010vf}
where exact agreement have been found.

\subsection{Integrable systems on the strip}

A relativistic integrable \emph{boundary} system in infinite volume
is defined on the negative \emph{half} line ($x\leq0$) only and characterized,
additionally to the dispersion relation and the two particle scattering
matrix, by the one particle \emph{reflection} matrix

\begin{center}
$\mathbb{R}(\theta)=R_{i}^{j}(\theta)$~~\includegraphics[height=1.5cm]{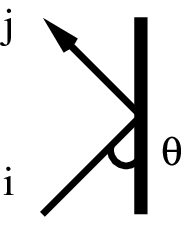}
\par\end{center}

\noindent which satisfies unitarity

\begin{center}
\includegraphics[height=2.5cm]{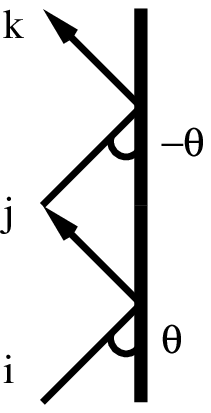}~~\includegraphics[height=2.5cm]{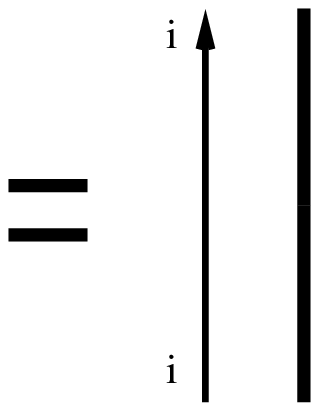}
\par\end{center}

\begin{equation}
R_{i}^{j}(\theta)R_{j}^{k}(-\theta)=\delta_{i}^{k}\label{eq:Runitarity}\end{equation}
 boundary crossing unitarity \cite{Ghoshal:1993tm} 

\begin{center}
\includegraphics[height=1.5cm]{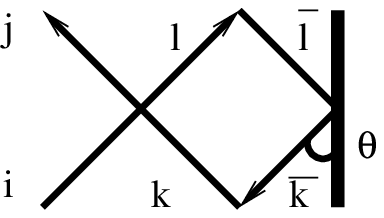}~~\includegraphics[height=1.5cm]{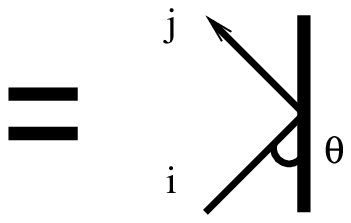}
\par\end{center}

\begin{equation}
S_{ik}^{lj}(2\theta)R_{\bar{k}}^{\bar{l}}(i\pi-\theta)=R_{i}^{j}(\theta)\label{eq:Bcrossunit}\end{equation}

\noindent and the boundary Yang-Baxter equation.

\begin{center}
\includegraphics[height=2.5cm]{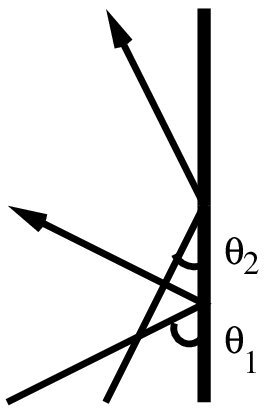}~~~~~~~~~~\includegraphics[height=2.5cm]{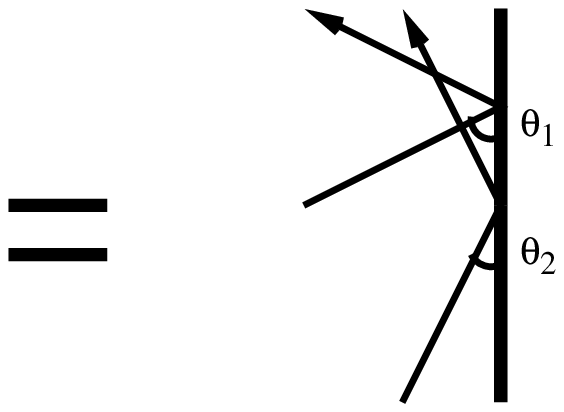}
\par\end{center}

\begin{equation}
\mathbb{S}(\theta_{1}-\theta_{2})\mathbb{R}(\theta_{1})\mathbb{S}(\theta_{1}+\theta_{2})\mathbb{R}(\theta_{2})=\mathbb{R}(\theta_{2})\mathbb{S}(\theta_{1}+\theta_{2})\mathbb{R}(\theta_{1})\mathbb{S}(\theta_{1}-\theta_{2})\label{eq:BYBE}\end{equation}

A finite volume boundary system of size $L$ has two boundaries with
left reflection factor $\mathbb{R}_{\alpha}$ and right reflection
factor $\mathbb{R}_{\beta}$. When a particle with positive rapidity
$\theta>0$ reflects back from the right boundary with reflection
factor $\mathbb{R}_{\beta}(\theta)$ it will reach the other boundary
with $-\theta$. It is a standard convention to denote the left reflection
factor of the particle with rapidity $-\theta$ by $\mathbb{R}_{\alpha}(\theta)$
as in this case $\mathbb{R}_{\alpha}(\theta)$ satisfies the same
equations (\ref{eq:Runitarity},\ref{eq:Bcrossunit},\ref{eq:BYBE})
as $\mathbb{R}_{\beta}(\theta)$ if the $S$-matrix is parity invariant,
which is usually the case. The energy levels of a multiparticle state
on the interval can be approximately described by the boundary BY
(BBY) equations:

\begin{center}
\includegraphics[height=3cm]{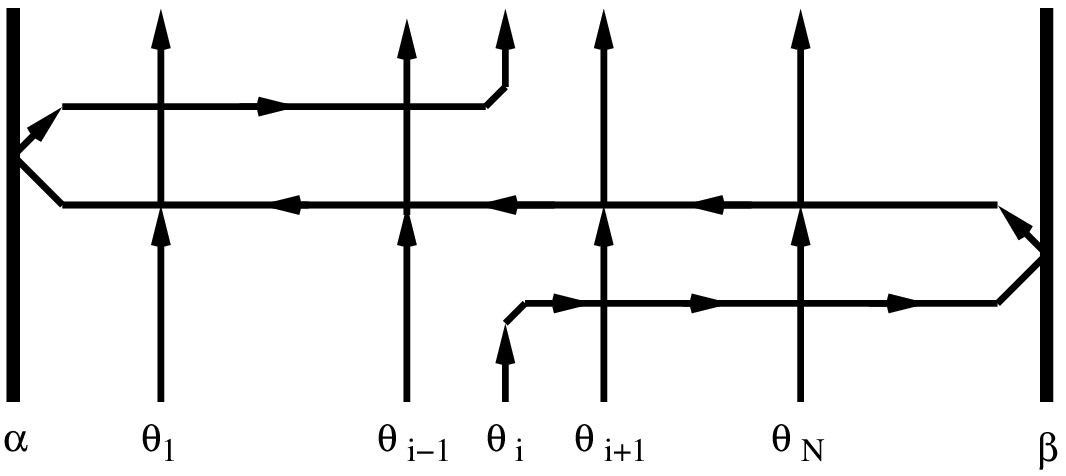}
\par\end{center}

\[
e^{2ip(\theta_{i})L}\prod_{j=i+1}^{N}\mathbb{S}(\theta_{i}-\theta_{j})\mathbb{R}_{\beta}(\theta_{i})\prod_{j=N}^{1}\mathbb{S}(\theta_{j}+\theta_{i})\mathbb{R}_{\alpha}(\theta_{i})\prod_{j=1}^{i-1}\mathbb{S}(\theta_{i}-\theta_{j})=\mathbb{I}\quad;\quad\theta_{i}>0\;;\; i=1,\dots,N\]
where similarly to the periodic case the product of reflection and
scattering matrices have to be diagonalized for each $i$. The energy
of the solution $E=\sum_{i=1}^{N}E(\theta_{i})$ contains all the
polynomial corrections in $L^{-1}$. 

Just as in the periodic case this can be nicely derived by introducing
the boundary analog of the transfer matrix, which is called the double-raw
transfer matrix%
\footnote{The concept of the double raw transfer matrix was introduced in integrable
boundary spin chains in \cite{Sklyanin:1988yz}. %
}

\begin{center}
\includegraphics[height=3cm]{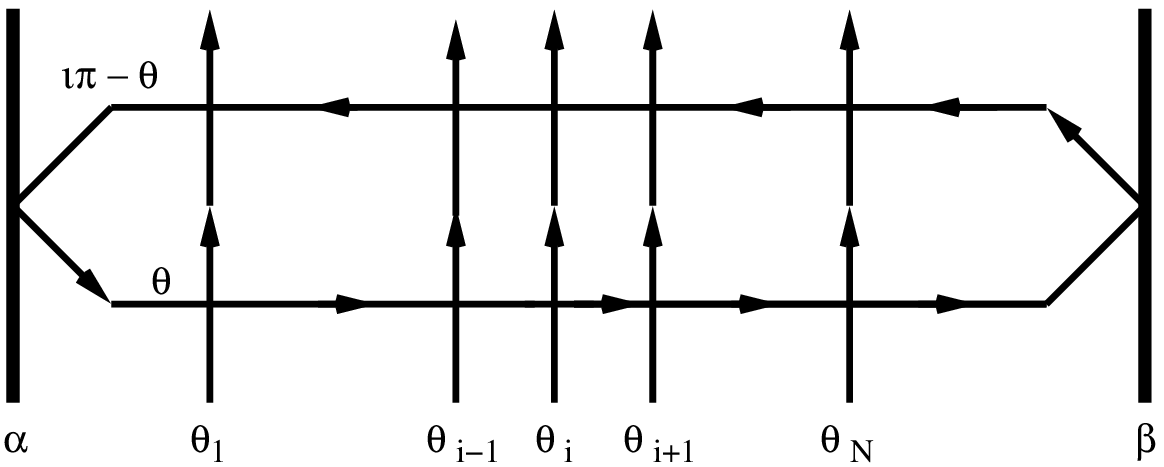}
\par\end{center}

\[
\mathbb{T}(\theta\vert\theta_{1},\dots,\theta_{N})=\mbox{Tr}\left(\prod_{j=1}^{N}\mathbb{S}(\theta-\theta_{j})\mathbb{R}_{\beta}(\theta)\prod_{j=N}^{1}\mathbb{S}(\theta_{j}+\theta)\mathbb{R}_{\alpha}^{c}(i\pi-\theta)\right)\]
where $\mathbb{R}^{c}$ is the charge conjugated reflection factor
$\mathbb{R}^{c}=\mathbb{CRC}^{-1}$. Observe that $\mathbb{R}_{\alpha}^{c}(i\pi-\theta)$
satisfies the consistency equations (\ref{eq:Runitarity},\ref{eq:Bcrossunit},\ref{eq:BYBE}),
whenever $\mathbb{R}_{\alpha}(\theta)$ satisfies them. The reason
why we have to use the reflection factor $\mathbb{R}_{\alpha}^{c}(i\pi-\theta)$
on the left boundary is that we would like to obtain the BBY equation.
In specifying the spectral parameter to any of the particles' rapidity
we have an extra scattering matrix $\mathbb{S}(2\theta_{i})$ which
should combine into $\mathbb{S}(2\theta_{i})\mathbb{R}_{\alpha}^{c}(i\pi-\theta_{i})=\mathbb{R}_{\alpha}(\theta)$.%
\footnote{A similar quantity was used in boundary kink theories to derive the
BBY equation from a double raw transfer matrix in \cite{Ahn:2000jd}.%
} 

Let us also mention that using the crossing symmetry of the scattering
matrix we might write the transfer matrix in an alternative way: \[
\mathbb{T}(\theta\vert\theta_{1},\dots,\theta_{N})=\mbox{Tr}\left(\prod_{j=1}^{N}\mathbb{S}(\theta-\theta_{j})\mathbb{R}_{\beta}(\theta)\prod_{j=N}^{1}\mathbb{S}^{c}(i\pi-\theta-\theta_{j})\mathbb{R}_{\alpha}^{c}(i\pi-\theta)\right)\]
where $\mathbb{S}^{c}$ is charge conjugated in the auxiliary space
only. This is the form of the double raw transfer matrix what is frequently
used in boundary lattice models, see \cite{Pearce:2000dv}. 

Due to the BYB and YB equations the transfer matrices commute for
different $\theta$s \cite{Sklyanin:1988yz}. The asymptotic $Y$-function
is defined after the diagonalization of this family of double-raw
transfer matrices. If the eigenvalue is denoted by $t(\theta\vert\theta_{1},\dots,\theta_{N})$
then the asymptotic $Y$-function is simply\[
Y_{as}(\theta\vert\theta_{1},\dots,\theta_{N})=e^{2ip(\theta)L}\, t(\theta\vert\theta_{1},\dots,\theta_{N})\]
One can check that, using the regularity of the scattering matrix,
the boundary crossing unitarity of the reflection factor and the bulk
YB equations, the BBY equation takes the same form as in the periodic
case: \[
Y_{as}(\theta_{i}\vert\theta_{1},\dots,\theta_{N})=-1\]
What is really nice in this formulation is that the exponentially
small finite size corrections can be described by exactly the same
formulas we had in the periodic case. The vacuum polarization effects
the energy of the $N$ particle state as \begin{equation}
\Delta E=-\int_{0}^{\infty}\frac{d\theta}{2\pi}\partial_{\theta}p(\theta)\, Y_{as}(\theta+i\frac{\pi}{2}\vert\theta_{1},\dots,\theta_{N})\label{eq:BLuscherrel}\end{equation}
 where due to the presence of the boundary we integrate only for positive
momentum particles. 

The leading finite size correction of the vacuum energy was derived
in \cite{Bajnok:2004tq} and checked against exact integral equations.
Evaluating $Y_{as}$ for the vacuum state \[
Y_{as}^{vac}(\theta+\frac{i\pi}{2})=\mbox{Tr}(\mathbb{R}_{\alpha}^{c}(\frac{i\pi}{2}-\theta)\mathbb{R}_{\beta}(\frac{i\pi}{2}+\theta))e^{-2mL\cosh\theta}\]
we can see that the generic formula (\ref{eq:BLuscherrel}) reproduces
the vacuum result \cite{Bajnok:2004tq}. In the case of excited states
the modification of the BBY equations are \[
\log(Y_{as}(\theta_{i}\vert\theta_{1},\dots,\theta_{N}))-\pi(2n+1)=-\partial_{\theta_{i}}\int_{0}^{\infty}\frac{d\theta}{2\pi}\, Y_{as}(\theta+i\frac{\pi}{2}\vert\theta_{1},\dots,\theta_{N})\]
We have checked that these formulas correctly reproduce the leading
exponential finite size corrections in the Lee-Yang and sinh-Gordon
and for certain states in the sine-Gordon models with Dirichlet boundary
conditions where exact integral equations were available \cite{Dorey:1997yg,Bajnok:2004tq,Bajnok:2007ep,Ahn:2003st}. 

In a nonrelativistic boundary theory these formulas have to be modified.
As the scattering matrix depends individually on its arguments the
boundary crossing equation is modified: \[
\mathbb{R}(u)=\mathbb{S}(u,-u)\mathbb{R}^{c}(\omega-u)\]
where $\mathbb{R}^{c}=\mathbb{CRC}^{-1}$. The BYBE takes the form
\[
\mathbb{S}(u_{1},u_{2})\mathbb{R}(u_{1})\mathbb{S}(u_{2},-u_{1})\mathbb{R}(u_{2})=\mathbb{R}(u_{2})\mathbb{S}(u_{2},-u_{1})\mathbb{R}(u_{1})\mathbb{S}(u_{1},u_{2})\]
Still one can define the double-raw transfer matrix \[
\mathbb{T}(u\vert u_{1},\dots,u_{N})=\mbox{Tr}\left(\prod_{j=1}^{N}\mathbb{S}(u,u_{j})\mathbb{R}_{\beta}(u)\prod_{j=N}^{1}\mathbb{S}(u_{j},-u)\mathbb{R}_{\alpha}^{c}(\omega-u)\right)\]
which commutes for different spectral parameters $u$. The crossing
symmetry of the bulk scattering matrix provides an equivalent formula
\[
\mathbb{T}(u\vert u_{1},\dots,u_{N})=\mbox{Tr}\left(\prod_{j=1}^{N}\mathbb{S}(u,u_{j})\mathbb{R}_{\beta}(u)\prod_{j=N}^{1}\mathbb{S}^{c}(\omega-u,u_{j})\mathbb{R}_{\alpha}^{c}(\omega-u)\right)\]
Using the eigenvalue of the transfer matrix one can define the asymptotic
$Y$- function \[
Y_{as}(u\vert u_{1},\dots,u_{N})=e^{2ip(u)L}t(u\vert u_{1},\dots,u_{N})\]
which can be used to describe the BBY equations \[
Y_{as}(u_{i}\vert u_{1},\dots,u_{N})=-1\]
The energy correction is expected to be \begin{equation}
\Delta E=-\int_{0}^{\infty}\frac{du}{2\pi}\partial_{u}\tilde{p}(u)\, Y_{as}(u+\frac{\omega}{2}\vert u_{1},\dots,u_{N})\label{eq:NonrelBLuscher}\end{equation}
For the vacuum state the correction reduces to $Y_{as}^{vac}(u+\frac{\omega}{2})=\mbox{Tr}(\mathbb{R}_{\alpha}^{c}(\frac{\omega}{2}-u)\mathbb{R}_{\beta}(\frac{\omega}{2}+u))e^{2iLp(u+\frac{\omega}{2})}$.
In this case there is no BBA equation to be modified. For a general
multiparticle state the modification of the BBA is conjectured to
be \[
\log Y_{as}(u_{i}\vert u_{1},\dots,u_{N})-\pi(2n+1)=\int_{0}^{\infty}\frac{du}{2\pi}\, t^{'}(u+\frac{\omega}{2}\vert u_{1},\dots,u_{N})e^{2ip(u+\frac{\omega}{2})L}\]
where $t^{'}(u\vert u_{1},\dots,u_{N})$ is nothing but the eigenvalue
of the operator

\[
\mbox{Tr}(\mathbb{S}(u,u_{1})\dots(\partial_{u}\mathbb{S}(u,u_{i}))\dots\mathbb{S}(u,u_{N})\mathbb{R}_{\beta}(u)\prod_{j=N}^{1}\mathbb{S}(u_{j},-u)\mathbb{R}_{\alpha}^{c}(\omega-u))\]
on the state under investigation and we supposed, similarly to the
periodic case, that the eigenstate of $\mathbb{T}$ is also an eigenstate
of this operator.

\section{Lüscher-type correction in AdS/CFT}

In this section we elaborate the previously conjectured finite size
energy correction formulas for a one particle state in the AdS/CFT
setting. In particular we focus on the $Y=0$ brane.

\subsubsection*{Lüscher-type correction }

In order to make connection to the general description we recall that
the dispersion relation of the $Q$ magnon bound states \[
E^{2}-16g^{2}\sin^{2}\frac{p}{2}=Q^{2}\]
can be uniformized on the torus with parameter $z$ in terms of the
Jacobi amplitudes as \cite{Janik:2006dc,Arutyunov:2007tc}: \[
p=2\mbox{ am}(z,k)\quad;\qquad E=Q\,\mbox{dn}(z,k)\quad;\qquad k=-16\frac{g^{2}}{Q^{2}}\]
The real period of the torus is $2\omega_{1}=4K(k)$ and the imaginary
period is $2\omega_{2}=4iK(1\lyxmathsym{\textminus}k)\lyxmathsym{\textminus}4K(k)$.
The crossing parameter in this theory is the half of the imaginary
period $\omega=\omega_{2}$ and $z$ plays the role of the generalized
rapidity, in terms of which the $S$-matrix satisfies, unitarity $\mathbb{S}(z_{1},z_{2})\mathbb{S}(z_{2},z_{1})=\mathbb{I}$,
crossing symmetry $\mathbb{S}^{c_{1}}(z_{1},z_{2})=\mathbb{S}(z_{2},z_{1}+\omega_{2})$
and the YBE \cite{Arutyunov:2008zt}. As we already mentioned, in
the AdS literature we use a different convention for the scattering
matrix, in which the ABA takes the form $e^{-ip_{j}L}\prod_{k}\mathbb{S}(p_{j},p_{k})=1$.
This means that instead of continuing to $u\to u+\frac{\omega}{2}$
we have to continue the result to $z\to z-\frac{\omega_{2}}{2}$. 

Taking this into account as a first application we take formula (\ref{eq:NonrelBLuscher})
and evaluate for the $\omega\to-\omega_{2}$ continuation and describe
the energy correction of the vacuum

\begin{center}
\includegraphics[height=3cm]{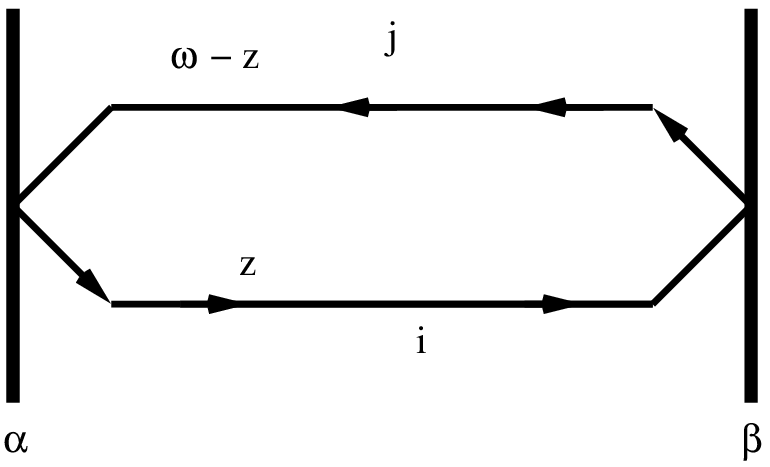}
\par\end{center}

\begin{equation}
\Delta E(L)=-\sum_{Q}\int_{0}^{\frac{\omega_{1}}{2}}\frac{dz}{2\pi}(\partial_{z}\tilde{p}_{Q}(z))\mathbb{R}_{i}^{j}(-\frac{\omega_{2}}{2}+z)\mathbb{C}_{j\bar{j}}\mathbb{R}_{\bar{i}}^{\bar{j}}(-\frac{\omega_{2}}{2}-z)\mathbb{C}^{i\bar{i}}e^{-2\tilde{\epsilon}_{Q}L}\label{eq:LuscherRz}\end{equation}
where we have to sum over the full infinite spectrum of the mirror
theory. This expression was evaluated in \cite{Correa:2009mz}. Let
us recall their results: first the reflection factors of the mirror
boundstates have to determined. According to the tensor product nature
of the mirror bound-states the reflection factor can be factorized
as \[
\mathbb{R}(z)=R_{0}(z)R(z)\otimes R(z)\]
Each $su(2\vert2)$ factor can be further decomposed with respect
to the unbroken $su(2)$ symmetry as $(Q+1;Q-1;Q;Q)$ where $Q$ is
the charge of the bound-state. Interestingly the unbroken $su(2\vert1)$
symmetry turns out to be restrictive enough to fix the matrix part
of the reflection factor completely:\[
R(z)=\mbox{diag}(\mathbb{I}_{Q+1};-\mathbb{I}_{Q-1};-e^{i\frac{p}{2}}\mathbb{I}_{Q},e^{-i\frac{p}{2}}\mathbb{I}_{Q})\]
This is in stark contrast to the case of physical bound-states which
transform under the totally symmetric representations. Indeed there
the unbroken $su(2\vert1)$ does not fix completely the matrix part
and higher symmetries as the Yangian have to be used. See \cite{Ahn:2010xa,MacKay:2010ey}
for an exhaustive analysis of the $Q=2$ case. The scalar part of
the bound-state reflection factor $R_{0}$ can be fixed from the fusion
principle and will be evaluated below. The nonzero matrix elements
of the fundamental charge conjugation matrix are $\mathbb{C}_{12}=-\mathbb{C}_{21}=-i$
and $\mathbb{C}_{34}=-\mathbb{C}_{43}=1$ from which it can easily
be extended to any representation. Evaluating the Lüscher correction
(\ref{eq:LuscherRz}) with the analytically continued bound-state
reflection factors gives vanishing result (independently of $R_{0}$)
which is consistent with the unbroken supersymmetry of the vacuum. 

Let us turn to the description of the finite size energy correction
of a one particle state. As the dispersion relation contain a factor
$g^{2}$ in front of $\sin^{2}\left(\frac{p}{2}\right)$, the weak
coupling expansion of the finite size corrections appearing in the
momentum quantization $p\to p+\delta p$ will be suppressed by one
order compared to the direct energy corrections. Consequently the
leading wrapping correction according to (\ref{eq:BLuscherrel}) turns
out to be 

\begin{center}
\includegraphics[height=3cm]{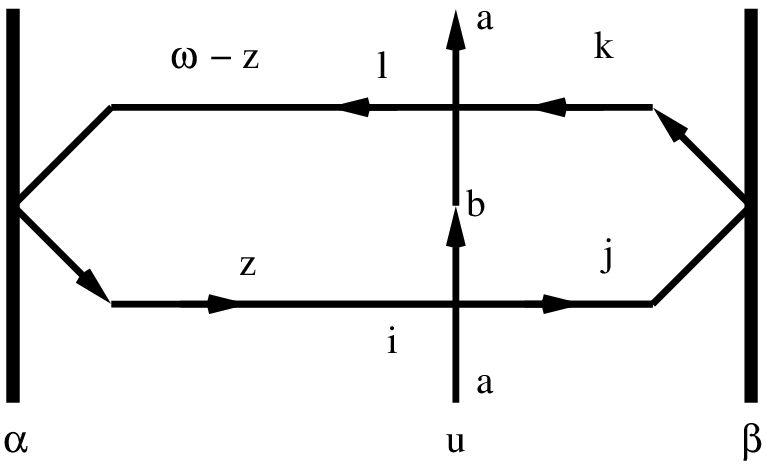}
\par\end{center}

\begin{equation}
\Delta E_{a}(L)=-\sum_{Q}\int_{0}^{\frac{\omega_{1}}{2}}\frac{dz}{2\pi}(\partial_{z}\tilde{p}_{Q}(z))\mathbb{S}_{ia}^{jb}(\frac{\omega}{2}+z,u)\mathbb{R}_{j}^{k}(\frac{\omega}{2}+z)\mathbb{S}_{lb}^{ka}(\frac{\omega}{2}-z,u)\mathbb{C}^{l\bar{l}}\mathbb{R}_{\bar{l}}^{\bar{i}}(\frac{\omega}{2}-z)\mathbb{C}_{\bar{i}i}e^{-2\tilde{\epsilon}_{Q}L}\label{eq:AdSLuscher}\end{equation}
where as we mentioned $\omega=-\omega_{2}$. An alternative formula
can be obtained by working directly in the mirror theory where the
same contribution can be depicted as

\begin{center}
\includegraphics[height=4cm]{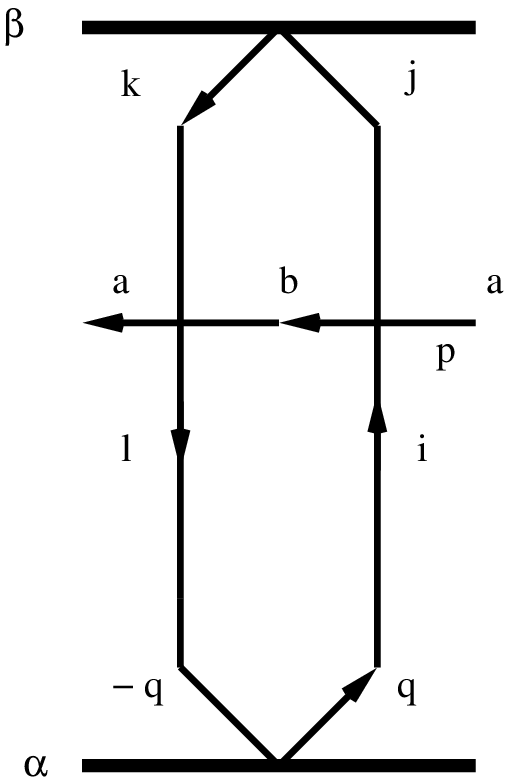}
\par\end{center}

\begin{equation}
\Delta E_{a}(L)=-\sum_{Q}\int_{0}^{\infty}\frac{dq}{2\pi}\mathbb{K}^{\bar{l}i}(q)\mathbb{S}_{ia}^{jb}(q,p)\bar{\mathbb{K}}_{j\bar{k}}(q)\mathbb{S}_{\bar{l}b}^{\bar{k}a}(-q,p)e^{-2\tilde{\epsilon}_{Q}L}\label{eq:KAdSLuscher}\end{equation}
and $a$ refers to the particle type whose energy correction we are
calculating. This expression contains the boundary state amplitudes,
$\mathbb{K}^{ij}(q)$ for each bound-state, which are related to the
reflection factors by analytical continuation $\mathbb{K}^{ij}(z)=\mathbb{C}^{i\bar{i}}\mathbb{R}_{\bar{i}}^{j}(\frac{\omega}{2}-z)$.
For convenience we write the formula in terms of the momentum of the
mirror particles $q=\tilde{p}_{Q}$.

Now we turn to the evaluation of the Lüscher formulas. Both the reflection
factors (boundary state amplitudes) and the scattering matrices can
be factorized according to the two to identical $su(2\vert2)$ {}``color''
factors: \[
\mathbb{S}(q,p)=S_{0}(q,p)\, S(q,p)\otimes S(q,p)\quad;\quad\mathbb{K}(q)=K_{0}(q)\, K(q)\otimes K(q)\]
In this decomposition the energy correction can be written as \begin{equation}
\Delta E(L)=-\sum_{Q}\int_{0}^{\infty}\frac{dq}{2\pi}K_{0}(q)S_{0}(q,p)\bar{K}_{0}(q)S_{0}(-q,p)\left[\mbox{Tr}\left(\bar{K}(q)S(q,p)K(q)S(-q,p)\right)\right]^{2}e^{-2\tilde{\epsilon}_{Q}L}\label{KLuscher}\end{equation}
where we calculate the Lüscher correction for a particle state labeled
by $a=(11)$ in (\ref{eq:KAdSLuscher}). This means that the operator
$\mbox{Tr}(\bar{K}(q)S(q,p)K(q)S(-q,p))$ acts diagonally on the one
particle states and we simply take its eigenvalue on the state labeled
by $a.$ We also note, that due to the particular definition of $\mathbb{K}$
and the charge conjugated $\bar{\mathbb{K}}$ (or the definition the
double raw transfer matrix containing $\mathbb{R}_{\alpha}^{c}(\omega-u)$)
we do not have to use supertrace. Alternatively, one can avoid the
charge conjugation on the right boundary and use the graded double
raw transfer matrix, see \cite{Murgan:2008fs} for a discussion of
this issue. 

As we are calculating the leading wrapping correction we expand all
functions in leading order in $g$. The expression $\tilde{\epsilon}_{Q}$
denotes the mirror energy of the charge $Q$ bound-state which in
the rapidity parametrization has the following weak coupling expansion:\[
e^{-\tilde{\epsilon}_{Q}}=\frac{4g^{2}}{q^{2}+Q^{2}}+O(g^{4})\]
 In the formula (\ref{eq:KAdSLuscher}) we have to sum also over all
polarization of the mirror bound-states. Let us focus on one copy
of the two $su(2\vert2)$ factors. The bound-states can be labeled
as follows: we decompose the $4Q$ dimensional completely antisymmetric
representation space as $4Q=(Q+1)+(Q-1)+Q+Q$ and parametrize the
sub-spaces in the superfield formalism \cite{Arutyunov:2007tc,Bajnok:2008bm}
by \begin{eqnarray*}
Q+1\longrightarrow\vert j\rangle^{1}= & \frac{1}{\sqrt{(Q-j)!j!}}w_{3}^{Q-j}w_{4}^{j} & \qquad;\qquad j=0,1,\dots,Q\\
Q-1\longrightarrow\vert j\rangle^{2}= & \frac{1}{\sqrt{(Q-2-j)!j!}}w_{3}^{Q-2-j}w_{4}^{j}\theta_{1}\theta_{2} & \qquad;\qquad j=0,1,\dots,Q-2\\
Q\longrightarrow\vert j\rangle^{3}= & \frac{1}{\sqrt{(Q-1-j)!j!}}w_{3}^{Q-1-j}w_{4}^{j}\theta_{1} & \qquad;\qquad j=0,1,\dots,Q-1\\
Q\longrightarrow\vert j\rangle^{4}= & \frac{1}{\sqrt{(Q-1-j)!j!}}w_{3}^{Q-1-j}w_{4}^{j}\theta_{2} & \qquad;\qquad j=0,1,\dots,Q-1\end{eqnarray*}
where we payed attention to the proper normalization of the states.
In this basis the boundary state amplitudes have the following nonzero
matrix elements \[
K_{j,Q-j}^{11}=(-1)^{j}\,;\quad K_{j,Q-2-j}^{22}=-(-1)^{j}\,;\quad K_{j,Q-1-j}^{34}=i(-1)^{j}e^{\frac{\tilde{\epsilon}_{Q}}{2}}\,;\quad K_{j,Q-1-j}^{43}=i(-1)^{j}e^{-\frac{\tilde{\epsilon}_{Q}}{2}}\]
 where the upper index refers to the subspace, while the lower labels
the state within. This form of the boundary state amplitudes follows
from requiring its vanishing under the unbroken $su(1\vert2)$ symmetry,
as we explicitly show in Appendix A. The conjugated boundary states
read as \[
\bar{K}_{j,Q-j}^{11}=(-1)^{j}\,;\quad\bar{K}_{j,Q-2-j}^{22}=-(-1)^{j}\,;\quad\bar{K}_{j,Q-1-j}^{43}=i(-1)^{j}e^{\frac{\tilde{\epsilon}_{Q}}{2}}\,;\quad\bar{K}_{j,Q-1-j}^{34}=i(-1)^{j}e^{-\frac{\tilde{\epsilon}_{Q}}{2}}\]
Thus basically a $3\leftrightarrow4$ change is made. 

The energy correction (\ref{eq:KAdSLuscher},\ref{KLuscher}) contains
also the scattering of the mirror bound-states with the fundamental
physical particle. To describe this scattering matrix we collect the
relevant coefficient from \cite{Bajnok:2008bm} in Appendix B. Using
the non-vanishing S-matrix elements from the Appendix of \cite{Bajnok:2008bm}
the contributions of the various subspaces can be written as:
\begin{itemize}
\item the $Q+1$ dimensional contribution \[
\sum_{j=0}^{Q}\bar{K}_{j,Q-j}^{11}\left[a_{5}^{5}(p,-q)K_{j,Q-j}^{11}a_{5}^{5}(p,q)-\frac{1}{2}a_{5}^{6}(p,-q)K_{j,Q-1-j}^{34}a_{2}^{3}(p,q)\right]\]

\item the $Q-1$ dimensional subspace contributes as \[
\sum_{j=0}^{Q-2}\bar{K}_{j,Q-2-j}^{22}\left[2a_{8}^{8}(p,-q)K_{j,Q-2-j}^{22}2a_{8}^{8}(p,q)+\frac{Q}{Q-1}a_{8}^{7}(p,-q)K_{j,Q-1-j}^{34}a_{4}^{3}(p,q)\right]\]

\item the $Q$ dimensional subspace with index $3$ as \[
\sum_{j=0}^{Q-1}\bar{K}_{j,Q-1-j}^{34}\left[a_{9}^{9}(p,-q)K_{j,Q-1-j}^{34}\frac{1}{2}(a_{9}^{9}(p,q)+a_{3}^{3}(p,q))\right]\]

\end{itemize}
finally the $Q$ dimensional subspace with index $4$ as \begin{eqnarray*}
\sum_{j=0}^{Q-1}\bar{K}_{j,Q-1-j}^{43}\biggr[\frac{1}{2}(a_{9}^{9}(p,-q)+a_{3}^{3}(p,-q))K_{j,Q-1-j}^{43}a_{9}^{9}(p,q)\,\,\,\,\,\,\,\,\,\,\,\,\,\,\,\,\,\,\,\,\,\,\,\,\,\,\,\,\,\,\,\,\,\,\,\,\,\,\,\\
+\frac{1}{2}(a_{9}^{9}(p,-q)-a_{3}^{3}(p,-q))K_{j,Q-1-j}^{34}\frac{1}{2}(a_{9}^{9}(p,q)-a_{3}^{3}(p,q))\\
\,\,\,\,\,\,\,\,\,\,\,\,\,\,\,\,\,\,\,\,\,\,\,\,\,\,-\frac{Q+1}{2Q}a_{2}^{3}(p,-q)K_{j,Q-j}^{11}a_{5}^{6}(p,q)+a_{4}^{3}(p,-q)K_{j,Q-2-j}^{22}a_{8}^{7}(p,q)\biggr]\end{eqnarray*}
where the $S_{ij}^{kl}$ $S$-matrix elements are obtained by multiplying the $a_m^n$ projector coefficients of \cite{Bajnok:2008bm} by $(-1)^{\epsilon_i \epsilon_j}$. 
We use the explicit expressions of $a_{i}^{j}(p,q)$ from Appendix
B together with the parametrization: \[
x^{\pm}(u)=\frac{2u\pm i}{4g}\left[1+\sqrt{1-\frac{16g^{2}}{(2u\pm i)^{2}}}\right]\quad;\qquad z^{\pm}(q)=\frac{q+iQ}{4g}\left[\sqrt{1+\frac{16g^{2}}{Q^{2}+q^{2}}}\pm1\right]\]
 and make the weak coupling expansion of each term. Here $u$ parametrizes
the physical momentum $p(u)$, and stands for the usual rapidity variable
of AdS/CFT and not for the generalized one $z$ (for which the crossing
equation is valid). The leading order contribution vanishes and for
the first non-vanishing contribution we obtain\[
\mbox{Tr}\left(\bar{K}(q)S(q,u)K(q)S(-q,u)\right)=-\frac{4Q(q^2+Q^2+1+4u^2)(q^2+Q^2-1-4u^2)}{(q^{2}+(-1+Q-2iu)^{2})(2u+i)^{3}(2u-i)}\]
The scalar part of the scattering matrix between the charge $Q$ mirror
bound-state and the fundamental physical particle according to \cite{Arutyunov:2009kf}
can be written as \[
S_{0}(q,u)=\frac{(x^{+}-z^{-})^{2}(-1+x^{-}z^{-})}{(x^{-}-z^{+})(x^{+}-z^{+})(-1+x^{+}z^{-})}\tilde{\Sigma}_{Q1}(q,u)\]
where $\tilde{\Sigma}$ contains the dressing phase and its weak coupling
expansion starts as $\tilde{\Sigma}_{Q1}(q,u)=1+g^{2}(\dots)$. Thus
the leading order expansion of the S-matrix scalar factor is \[
S_{0}(q,u)=\frac{(2u+i)^{2}(2u-q+i(Q-1))}{(2u-q-i(Q+1))(2u-q-i(Q-1))(2u-q+i(Q+1))}+O(g^{2})\]
 The scalar part of the boundary state amplitude for a charged $Q$
mirror boundstate can be analytically continued from the boundstate
reflection factor, whose scalar part can be calculated by the bootstrap
principle \cite{Ahn:2007bq}. The charge $Q$ bound-state composed
of elementary magnons as $x=(x_{1},\dots,x_{Q})$, such that $x^{-}=x_{1}^{-}$
and $x_{Q}^{+}=x^{+}$and the bound-state condition is also satisfied
$x_{i}^{+}=x_{i+1}^{-}$. Thus the full scalar factor as the product
of the elementary scalar factors turns out to be : \begin{equation}
R_{0}^{Q}(x)=\prod_{i=1}^{Q}R_{0}^{1}(x_{i}^{\pm})\prod_{i<j}S_{0}^{11}(x_{i}^{\pm},-x_{j}^{\mp})\end{equation}
where $R_{0}^{1}(x^{\pm})$ denotes the scalar factor of a fundamental
particle, (\ref{eq:R0p}), while $S_{0}^{11}(x_{1},x_{2})$ denotes
the scalar factor of their scattering matrices. The combination appearing
in $K_{0}\bar{K}_{0}$ can be easily calculated following \cite{Correa:2009mz}
and one finds that \[
\bar{K}_{0}(q)K_{0}(q)=\frac{4(1+z^{+}z^{-})^{2}}{(z^{+}+\frac{1}{z^{+}})(z^{-}+\frac{1}{z^{-}})(z^{-}+z^{+})^{2}}=\frac{256q^{2}g^{4}}{(q^{2}+Q^{2})^{3}}+\dots\]
Putting together in eq. (\ref{KLuscher}) the contributions of the
matrix part and the scalar parts of the $S$-matrices and the boundary
states one obtains that the leading correction is proportional to
$g^{4(L+1)}$ in case of a strip with width $L$. Evaluating the correction
(\ref{KLuscher}) we must set the momentum (or rapidity) of the fundamental
particle equal to the value(s) allowed by the BBY equation for the
particular $L$. Thus for $J=3$ $(L=2)$ we have to use $p=\frac{\pi}{2}$.
Evaluating the formula we obtain\[
\Delta E=192g^{12}(4\zeta(5)-7\zeta(9)) \]
The integrand satisfied a very nontrivial consistency relation, namely
the contribution of the dynamical poles vanished when we summed them
over the boundstate spectrum ($Q$). The transcendentality property
of $\Delta E$ is what is expected from a six loop gauge theory
calculation: the maximal transcendentality is $2(\#\mbox{loops})-3$,
as it happened for the Konishi operator at four and five loops. It
is interesting to compare our result to the analogous result for the
anomalous dimension of the Konishi operator in the periodic theory:
first the wrapping correction in the strip geometry consists of the
linear combinations of $\zeta$ functions only without the additional
integer term, and second it seems to appear in relatively higher orders
in $g$. This is a generic feature of the boundary finite size corrections
as they start at $e^{-2mL}$ compared to the periodic case which starts
at $e^{-mL}$. There is one exceptional case, namely when the boundary
reflection factor admits a kinematical pole at $q=0$, since then
the boundary correction starts also at $e^{-mL}.$ This phenomena
does not appear for the $Y=0$ brane, but it indeed happens for $Z=0$
\cite{Correa:2009mz}.

\section{Conclusions}

In this paper we have proposed formulas to describe the leading Lüscher-type
finite size energy correction for multiparticle states on the strip.
By this we generalized two results in the literature. On one hand
we generalized the multiparticle finite size energy corrections from
the periodic (cylinder) \cite{Bajnok:2008bm} to the boundary (strip)
setting. On the other we extended the boundary finite size correction
from the vacuum state \cite{Bajnok:2004tq,Bajnok:2006dn,Correa:2009mz}
to excited multiparticle states. We then evaluated the proposed formulas
for a single particle excitation reflecting diagonally on the $Y=0$
brane and determined the wrapping contribution to the anomalous dimension
of the simplest determinant operator of the form $O_{Y}(ZYZ^{J})$. 

The calculation, after internal consistency checks, resulted in sums
of $\zeta$-functions, which are also expected from a gauge theory
calculation. Nevertheless a stringent consistency check could be obtained
by calculating the wrapping contribution directly from Feynman diagrams
on the gauge theory side. This can be done after identifying the wrapping
type diagrams, analogously to the case what was developed for single
trace operators \cite{Sieg:2005kd}. 

In this paper we calculated the finite size correction to the energy
of the simplest one particle state. Clearly the approach is quite
general and can be applied, in principle, to any multiparticle state
over the $Y=0$ ground state, although it can be cumbersome to collect
the various S-matrix elements. A bit more sophisticated approach can
be based on the Y-system. In \cite{Gromov:2009tv} $Y$-system type
functional relations was proposed to describe the spectrum of planar
AdS/CFT and, in the same time, its asymptotic solution was expressed
in terms of the transfer matrix eigenvalues of the ABA. We expect
that the same $Y$-system describes the boundary AdS/CFT, too, and
that the corresponding asymptotical solution can be expressed also
in terms of the eigenvalues of the double raw transfer matrices. A
work is in progress in this direction. 

Although the conjectured $Y$-system could describe all the excited
states, it is useful only if its analytical structure is completely
understood. A derivation based on the boundary Thermodynamical Bethe
Ansatz would provide not only the rigorous establishment of the $Y$-system,
but also present the needed analytical structure. 

Recently there is a growing interest in the less supersymmetric, $\beta$-deformed,
version of the AdS/CFT correspondence \cite{Ahn:2010yv,Beisert:2005if,Arutyunov:2010gu,Gromov:2010dy,Ahn:2010ws}.
We believe that our boundary formulation can be extended to this realm,
too.

\subsection*{Acknowledgments}

We thank Changrim Ahn, Diego Correa, Romuald Janik, Rafael Nepomechie,
Christoph Sieg, Ryo Suzuki, Charles Young for the useful discussions.
Part of this work was performed at the workshop {}``Finite-Size Technology
in Low-Dimensional Quantum Systems V'' (Benasque). The work was supported
by a Bolyai Scholarship, and by OTKA K81461 and K60040.

\appendix

\section{Mirror boundary state}

In this Appendix we show that the matrix structure of the boundary
state amplitudes for each $Q$ (mirror) bound states can be obtained
by requiring the boundary state to be a singlet under the unbroken
$su(1\vert2)$ symmetry. The boundary state can be written as \[
\vert B\rangle=\sum\limits _{Q=1}^{\infty}\vert B_{Q}\rangle,\qquad\vert B_{Q}\rangle=\sum\limits _{a,b=1}^{4}\sum\limits _{ij}K_{ij}^{ab}\vert i\rangle_{(-q)}^{a}\otimes\vert j\rangle_{(q)}^{b}\]
 where $a,b$ are running over the the four subspaces of the $4Q$
dimensional representation space of the mirror $Q$ bound state and
$\vert j\rangle_{(q)}^{b}$ represent (the superfields description
of) the states belonging to them as described in the main body of
the paper. (The subscript on these symbols indicate the momentum of
the bound state). The bosonic generators of the unbroken symmetry
can be described in terms of the fermionic $\theta_{1}$ $\theta_{2}$
and bosonic $w_{3}$, $w_{4}$ parameters as \[
L_{1}^{1}=\frac{1}{2}(\theta_{1}\frac{\partial}{\partial\theta_{1}}-\theta_{2}\frac{\partial}{\partial\theta_{2}}),\quad R_{\alpha}^{\beta}=w_{\alpha}\frac{\partial}{\partial w_{\beta}}-\frac{1}{2}\delta_{\alpha}^{\beta}w_{\gamma}\frac{\partial}{\partial w_{\gamma}},\quad\alpha,\,\beta=3,\,4\]
 The requirements $L_{1}^{1}\vert B_{Q}\rangle=0$, $R_{\alpha}^{\beta}\vert B_{Q}\rangle=0$
restrict the form of $\vert B_{Q}\rangle$ as \[
\vert B_{Q}\rangle=K^{11}\sum\limits _{j=0}^{Q}(-1)^{j}\vert j\rangle_{(-q)}^{1}\otimes\vert Q-j\rangle_{(q)}^{1}+K^{22}\sum\limits _{j=0}^{Q-2}(-1)^{j}\vert j\rangle_{(-q)}^{2}\otimes\vert Q-2-j\rangle_{(q)}^{2}\]
 \[
\ \ \quad+K^{34}\sum\limits _{j=0}^{Q-1}(-1)^{j}\vert j\rangle_{(-q)}^{3}\otimes\vert Q-1-j\rangle_{(q)}^{4}+K^{43}\sum\limits _{j=0}^{Q-1}(-1)^{j}\vert j\rangle_{(-q)}^{4}\otimes\vert Q-1-j\rangle_{(q)}^{3}\]
 To impose also $Q_{\alpha}^{1}\vert B_{Q}\rangle=0$ we need not
only the explicit form of the supersymmetry generators in the superfield
formalism \[
Q_{\alpha}^{1}=aw_{\alpha}\frac{\partial}{\partial\theta_{1}}+b\epsilon^{12}\epsilon_{\alpha\beta}\theta_{2}\frac{\partial}{\partial w_{\beta}}\]
 but also the fact, that they have a non trivial coproduct \cite{Arutyunov:2007tc}:
\[
Q_{\alpha}^{1}(\vert j\rangle_{(-q)}^{b}\otimes\vert l\rangle_{(q)}^{b})=e^{\epsilon_{Q}(q)/4}(Q_{\alpha}^{1}\vert j\rangle_{(-q)}^{b})\otimes\vert l\rangle_{(q)}^{b}+e^{-\epsilon_{Q}(q)/4}\vert j\rangle_{(-q)}^{b}\otimes(Q_{\alpha}^{1}\vert l\rangle_{(q)}^{b})\]
 Here $a,b$ are the $q$ dependent coefficients, obtained by the
$x^{\pm}(p)\mapsto z^{\pm}(q)$ analytic continuation from the magnon
channel \[
a(q)=\sqrt{\frac{g}{Q}}\eta(q),\quad b(q)=\sqrt{\frac{g}{Q}}\frac{i}{\eta(q)}(\frac{z^{+}}{z^{-}}-1),\quad\eta(q)=e^{\epsilon_{Q}(q)/4}\sqrt{i(z^{-}-z^{+})}\]
 and we also exploited that $\epsilon_{Q}(q)=\epsilon_{Q}(-q)$. The
requirement $Q_{\alpha}^{1}\vert B_{Q}\rangle=0$ leads to a (compatible)
homogeneous linear system of equations for the remaining $K^{ab}$.
The solution can be written in terms of the undetermined $K^{11}$
as \[
K^{22}=-K^{11},\quad K^{34}=-iK^{11}e^{-\epsilon_{Q}(q)/2},\quad K^{43}=-iK^{11}e^{\epsilon_{Q}(q)/2}\]
If we,  instead of demanding the conservation of $Q_{\alpha}^{1} $,  impose the $Q_{\alpha}^{2}\vert B_{Q}\rangle=0$
requirement the analogous calculation yields
\[
K^{22}=-K^{11},\quad K^{34}=iK^{11}e^{\epsilon_{Q}(q)/2},\quad K^{43}=iK^{11}e^{-\epsilon_{Q}(q)/2}\]
This solution is equivalent to the one obtained by analytical
continuation from the diagonal reflection matrix $R(z)$, therefore we use this boundary state 
to calculate the Lüscher correction.

\section{Scattering matrix coefficients}

Here we collect the explicit form of the scattering matrix elements
we used to calculate the finite size corrections. 

\[
a_{5}^{5}=\frac{x_{1}^{+}-x_{2}^{+}}{x_{1}^{+}-x_{2}^{-}}\frac{\tilde{\eta}_{1}}{\eta_{1}}\]

\[
a_{6}^{5}=\sqrt{Q}\frac{(x_{2}^{+}-x_{2}^{-})}{(x_{1}^{+}-x_{2}^{-})}\frac{\tilde{\eta}_{1}}{\eta_{2}}\]
\[
a_{9}^{9}=\frac{x_{1}^{-}-x_{2}^{+}}{x_{1}^{+}-x_{2}^{-}}\frac{\tilde{\eta}_{1}}{\eta_{1}}\frac{\tilde{\eta}_{2}}{\eta_{2}}\]
\[
a_{8}^{8}=\frac{1}{2}\frac{x_{1}^{-}(x_{1}^{-}-x_{2}^{+})(1-x_{1}^{+}x_{2}^{-})}{x_{1}^{+}(x_{1}^{+}-x_{2}^{-})(1-x_{1}^{-}x_{2}^{-})}\frac{\tilde{\eta}_{1}}{\eta_{1}}\frac{\tilde{\eta}_{2}^{2}}{\eta_{2}^{2}}\]
\[
a_{7}^{8}=-\frac{i}{\sqrt{Q}}\frac{x_{1}^{-}x_{2}^{-}(x_{1}^{-}-x_{2}^{+})}{x_{1}^{+}x_{2}^{+}(x_{1}^{+}-x_{2}^{-})(1-x_{1}^{-}x_{2}^{-})}\frac{\tilde{\eta}_{1}\tilde{\eta}_{2}^{2}}{\eta_{2}}\]
\[
a_{3}^{2}=-\frac{2i}{\sqrt{Q}}\frac{(x_{1}^{-}-x_{1}^{+})(x_{2}^{-}-x_{2}^{+})(x_{1}^{+}-x_{2}^{+})}{(x_{1}^{+}-x_{2}^{-})(1-x_{1}^{-}x_{2}^{-})\eta_{1}\eta_{2}}\]
\[
a_{4}^{3}=\frac{Q-1}{\sqrt{Q}}\frac{x_{1}^{-}(x_{2}^{+}-x_{2}^{-})(1-x_{1}^{+}x_{2}^{-})}{x_{1}^{+}(x_{1}^{+}-x_{2}^{-})(1-x_{1}^{-}x_{2}^{-})}\frac{\tilde{\eta}_{1}\tilde{\eta}_{2}}{\eta_{2}^{2}}\]
\[
a_{3}^{3}=-\frac{(-x_{1}^{-}x_{1}^{+}(1+x_{1}^{-}x_{2}^{-}-2x_{1}^{+}x_{2}^{-})-(x_{1}^{+}+x_{1}^{-}(-2+x_{1}^{+}x_{2}^{-}))x_{2}^{+})}{x_{1}^{+}(x_{1}^{+}-x_{2}^{-})(1-x_{1}^{-}x_{2}^{-})}\frac{\tilde{\eta}_{1}}{\eta_{1}}\frac{\tilde{\eta}_{2}}{\eta_{2}}\]

where the following phase factors have been chosen:\[
\tilde{\eta}_{1}=e^{i\frac{p_{1}}{4}}\sqrt{i(x_{1}^{-}-x_{1}^{+})}\quad;\qquad\eta_{1}=e^{i\frac{p_{2}}{2}}\tilde{\eta}_{1}\quad;\qquad\eta_{2}=e^{i\frac{p_{2}}{4}}\sqrt{i(x_{2}^{-}-x_{2}^{+})}\quad;\qquad\tilde{\eta}_{2}=e^{i\frac{p_{1}}{2}}\eta_{2}\]

\section{Complete solution of the BBY}

In this appendix we present the complete weak coupling solution of
the BBY (\ref{eq:bby}) up to the order of $g^{4}$. To describe them
we introduce $J=L+1$. Since $p$ must be in the range $0<p<\pi$,
its allowed values are as follows: 

\begin{center}
\begin{tabular}{|c|c|c|}
\hline 
magnon labels  & allowed $p$-s  & nature \tabularnewline
\hline 
(33)\,(34)\,(43)\,(44)\,(12)\,(21)  & $p_{n}=n\frac{\pi}{J}$, $\quad n=1,\dots J-1$  & bosonic\tabularnewline
\hline 
(11)  & $p_{n}=n\frac{\pi}{J-1}$, $\quad n=1,\dots J-2$  & bosonic \tabularnewline
\hline 
(22)  & $p_{n}=n\frac{\pi}{J+1}$, $\quad n=1,\dots J$  & bosonic \tabularnewline
\hline 
(13)\,(14)\,(31)\,(41)  & $p_{n}=\frac{2n\pi}{2J-1}$, $\quad n=1,\dots J-1$  & fermionic \tabularnewline
\hline 
(23)\,(24)\,(32)\,(42)  & $p_{n}=\frac{2n\pi}{2J+1}$, $\quad n=1,\dots J$  & fermionic \tabularnewline
\hline
\end{tabular}
\par\end{center}

\noindent (In writing the entries of the table we also exploited that
$J$ is an integer).

The different allowed momenta for various different magnon labels
indicate that the presence of the two boundaries splits the 16-fold
degeneracy of the bulk magnon states. A particularly interesting aspect
of these allowed $p$ values is to consider the difference between
the sum of bosonic and fermionic energies $E_{B}-E_{F}$ where \[
E_{B}=6\sum\limits _{n=1}^{J-1}\sqrt{1+16g^{2}\sin^{2}(\frac{n\pi}{2J})}+\sum\limits _{n=1}^{J-2}\sqrt{1+16g^{2}\sin^{2}(\frac{n\pi}{2(J-1)})}+\sum\limits _{n=1}^{J}\sqrt{1+16g^{2}\sin^{2}(\frac{n\pi}{2(J+1)})}\]
 and \[
E_{F}=4\bigl(\sum\limits _{n=1}^{J-1}\sqrt{1+16g^{2}\sin^{2}(\frac{n\pi}{2J-1})}+\sum\limits _{n=1}^{J}\sqrt{1+16g^{2}\sin^{2}(\frac{n\pi}{2J+1})}\bigr)\]
 since this difference may be thought of as the sum of the (single
magnon) vacuum fluctuations (zero mode sum) around the ground state
(i.e. two boundaries with no particle between them). The interesting
observation is that adding to this half of the energies of the $p=\pi$
bosonic modes $E_{Z}/2=4\sqrt{1+16g^{2}}$ we get precisely zero \[
E_{B}+E_{Z}/2-E_{F}=0.\]
 We checked this analytically up to $g^{8}$ for a number of (integer)
$J$-s and also numerically for some randomly chosen $g$-s. The vanishing
of this sum may be consistent with the ground state preserving one
supersymmetry. It would be interesting to perform an analogous calculation
for the $Q=2$ bound-state particles \cite{Ahn:2010xa}.


\begin{thebibliography}{10}
\expandafter\ifx\csname url\endcsname\relax
  \def\url#1{\texttt{#1}}\fi
\expandafter\ifx\csname urlprefix\endcsname\relax\def\urlprefix{URL }\fi
\expandafter\ifx\csname href\endcsname\relax
  \def\href#1#2{#2} \def\path#1{#1}\fi

\bibitem{Maldacena:1997re}
J.~M. Maldacena, {The large N limit of superconformal field theories and
  supergravity}, Adv. Theor. Math. Phys. 2 (1998) 231--252.
\newblock \href {http://arxiv.org/abs/hep-th/9711200}
  {\path{arXiv:hep-th/9711200}}.

\bibitem{Gubser:1998bc}
S.~S. Gubser, I.~R. Klebanov, A.~M. Polyakov, {Gauge theory correlators from
  non-critical string theory}, Phys. Lett. B428 (1998) 105--114.
\newblock \href {http://arxiv.org/abs/hep-th/9802109}
  {\path{arXiv:hep-th/9802109}}, \href
  {http://dx.doi.org/10.1016/S0370-2693(98)00377-3}
  {\path{doi:10.1016/S0370-2693(98)00377-3}}.

\bibitem{Witten:1998qj}
E.~Witten, {Anti-de Sitter space and holography}, Adv. Theor. Math. Phys. 2
  (1998) 253--291.
\newblock \href {http://arxiv.org/abs/hep-th/9802150}
  {\path{arXiv:hep-th/9802150}}.

\bibitem{Minahan:2002ve}
J.~A. Minahan, K.~Zarembo, {The Bethe-ansatz for N = 4 super Yang-Mills}, JHEP
  03 (2003) 013.
\newblock \href {http://arxiv.org/abs/hep-th/0212208}
  {\path{arXiv:hep-th/0212208}}.

\bibitem{Beisert:2003tq}
N.~Beisert, C.~Kristjansen, M.~Staudacher, {The dilatation operator of N = 4
  super Yang-Mills theory}, Nucl. Phys. B664 (2003) 131--184.
\newblock \href {http://arxiv.org/abs/hep-th/0303060}
  {\path{arXiv:hep-th/0303060}}, \href
  {http://dx.doi.org/10.1016/S0550-3213(03)00406-1}
  {\path{doi:10.1016/S0550-3213(03)00406-1}}.

\bibitem{Bena:2003wd}
I.~Bena, J.~Polchinski, R.~Roiban, {Hidden symmetries of the AdS(5) x S**5
  superstring}, Phys. Rev. D69 (2004) 046002.
\newblock \href {http://arxiv.org/abs/hep-th/0305116}
  {\path{arXiv:hep-th/0305116}}, \href
  {http://dx.doi.org/10.1103/PhysRevD.69.046002}
  {\path{doi:10.1103/PhysRevD.69.046002}}.

\bibitem{Kazakov:2004qf}
V.~A. Kazakov, A.~Marshakov, J.~A. Minahan, K.~Zarembo, {Classical / quantum
  integrability in AdS/CFT}, JHEP 05 (2004) 024.
\newblock \href {http://arxiv.org/abs/hep-th/0402207}
  {\path{arXiv:hep-th/0402207}}.

\bibitem{Arutyunov:2004vx}
G.~Arutyunov, S.~Frolov, M.~Staudacher, {Bethe ansatz for quantum strings},
  JHEP 10 (2004) 016.
\newblock \href {http://arxiv.org/abs/hep-th/0406256}
  {\path{arXiv:hep-th/0406256}}, \href
  {http://dx.doi.org/10.1088/1126-6708/2004/10/016}
  {\path{doi:10.1088/1126-6708/2004/10/016}}.

\bibitem{Staudacher:2004tk}
M.~Staudacher, {The factorized S-matrix of CFT/AdS}, JHEP 05 (2005) 054.
\newblock \href {http://arxiv.org/abs/hep-th/0412188}
  {\path{arXiv:hep-th/0412188}}.

\bibitem{Beisert:2005tm}
N.~Beisert, {The su(2|2) dynamic S-matrix}, Adv. Theor. Math. Phys. 12 (2008)
  945.
\newblock \href {http://arxiv.org/abs/hep-th/0511082}
  {\path{arXiv:hep-th/0511082}}.

\bibitem{Beisert:2005fw}
N.~Beisert, M.~Staudacher, {Long-range PSU(2,2|4) Bethe ansaetze for gauge
  theory and strings}, Nucl. Phys. B727 (2005) 1--62.
\newblock \href {http://arxiv.org/abs/hep-th/0504190}
  {\path{arXiv:hep-th/0504190}}, \href
  {http://dx.doi.org/10.1016/j.nuclphysb.2005.06.038}
  {\path{doi:10.1016/j.nuclphysb.2005.06.038}}.

\bibitem{Beisert:2004hm}
N.~Beisert, V.~Dippel, M.~Staudacher, {A novel long range spin chain and planar
  N = 4 super Yang- Mills}, JHEP 07 (2004) 075.
\newblock \href {http://arxiv.org/abs/hep-th/0405001}
  {\path{arXiv:hep-th/0405001}}.

\bibitem{Luscher:1986pf}
M.~Luscher, {Volume Dependence of the Energy Spectrum in Massive Quantum Field
  Theories. 2. Scattering States}, Commun. Math. Phys. 105 (1986) 153--188.
\newblock \href {http://dx.doi.org/10.1007/BF01211097}
  {\path{doi:10.1007/BF01211097}}.

\bibitem{Ambjorn:2005wa}
J.~Ambjorn, R.~A. Janik, C.~Kristjansen, {Wrapping interactions and a new
  source of corrections to the spin-chain / string duality}, Nucl. Phys. B736
  (2006) 288--301.
\newblock \href {http://arxiv.org/abs/hep-th/0510171}
  {\path{arXiv:hep-th/0510171}}, \href
  {http://dx.doi.org/10.1016/j.nuclphysb.2005.12.007}
  {\path{doi:10.1016/j.nuclphysb.2005.12.007}}.

\bibitem{Bajnok:2008bm}
Z.~Bajnok, R.~A. Janik, {Four-loop perturbative Konishi from strings and finite
  size effects for multiparticle states}, Nucl. Phys. B807 (2009) 625--650.
\newblock \href {http://arxiv.org/abs/0807.0399} {\path{arXiv:0807.0399}},
  \href {http://dx.doi.org/10.1016/j.nuclphysb.2008.08.020}
  {\path{doi:10.1016/j.nuclphysb.2008.08.020}}.

\bibitem{Bajnok:2009vm}
Z.~Bajnok, A.~Hegedus, R.~A. Janik, T.~Lukowski, {Five loop Konishi from
  AdS/CFT, }\href {http://arxiv.org/abs/0906.4062} {\path{arXiv:0906.4062}}.

\bibitem{Fiamberti:2007rj}
F.~Fiamberti, A.~Santambrogio, C.~Sieg, D.~Zanon, {Wrapping at four loops in
  N=4 SYM}, Phys. Lett. B666 (2008) 100--105.
\newblock \href {http://arxiv.org/abs/0712.3522} {\path{arXiv:0712.3522}},
  \href {http://dx.doi.org/10.1016/j.physletb.2008.06.061}
  {\path{doi:10.1016/j.physletb.2008.06.061}}.

\bibitem{Fiamberti:2008sh}
F.~Fiamberti, A.~Santambrogio, C.~Sieg, D.~Zanon, {Anomalous dimension with
  wrapping at four loops in N=4 SYM}, Nucl. Phys. B805 (2008) 231--266.
\newblock \href {http://arxiv.org/abs/0806.2095} {\path{arXiv:0806.2095}},
  \href {http://dx.doi.org/10.1016/j.nuclphysb.2008.07.014}
  {\path{doi:10.1016/j.nuclphysb.2008.07.014}}.

\bibitem{Velizhanin:2008jd}
V.~N. Velizhanin, {The Four-Loop Konishi in N=4 SYM, }\href
  {http://arxiv.org/abs/0808.3832} {\path{arXiv:0808.3832}}.

\bibitem{Bajnok:2008qj}
Z.~Bajnok, R.~A. Janik, T.~Lukowski, {Four loop twist two, BFKL, wrapping and
  strings}, Nucl. Phys. B816 (2009) 376--398.
\newblock \href {http://arxiv.org/abs/0811.4448} {\path{arXiv:0811.4448}},
  \href {http://dx.doi.org/10.1016/j.nuclphysb.2009.02.005}
  {\path{doi:10.1016/j.nuclphysb.2009.02.005}}.

\bibitem{Beccaria:2009eq}
M.~Beccaria, V.~Forini, T.~Lukowski, S.~Zieme, {Twist-three at five loops,
  Bethe Ansatz and wrapping}, JHEP 03 (2009) 129.
\newblock \href {http://arxiv.org/abs/0901.4864} {\path{arXiv:0901.4864}},
  \href {http://dx.doi.org/10.1088/1126-6708/2009/03/129}
  {\path{doi:10.1088/1126-6708/2009/03/129}}.

\bibitem{Beccaria:2009hg}
M.~Beccaria, G.~F. De~Angelis, {On the wrapping correction to single magnon
  energy in twisted N=4 SYM}, Int. J. Mod. Phys. A24 (2009) 5803--5817.
\newblock \href {http://arxiv.org/abs/0903.0778} {\path{arXiv:0903.0778}},
  \href {http://dx.doi.org/10.1142/S0217751X09047375}
  {\path{doi:10.1142/S0217751X09047375}}.

\bibitem{Beccaria:2009vt}
M.~Beccaria, V.~Forini, {Four loop reciprocity of twist two operators in N=4
  SYM}, JHEP 03 (2009) 111.
\newblock \href {http://arxiv.org/abs/0901.1256} {\path{arXiv:0901.1256}},
  \href {http://dx.doi.org/10.1088/1126-6708/2009/03/111}
  {\path{doi:10.1088/1126-6708/2009/03/111}}.

\bibitem{Lukowski:2009ce}
T.~Lukowski, A.~Rej, V.~N. Velizhanin, {Five-Loop Anomalous Dimension of
  Twist-Two Operators, }\href {http://arxiv.org/abs/0912.1624}
  {\path{arXiv:0912.1624}}.

\bibitem{Velizhanin:2010cm}
V.~N. Velizhanin, {Six-Loop Anomalous Dimension of Twist-Three Operators in N=4
  SYM}\href {http://arxiv.org/abs/1003.4717} {\path{arXiv:1003.4717}}.

\bibitem{Gromov:2009tv}
N.~Gromov, V.~Kazakov, P.~Vieira, {Exact Spectrum of Anomalous Dimensions of
  Planar N=4 Supersymmetric Yang-Mills Theory}, Phys. Rev. Lett. 103 (2009)
  131601.
\newblock \href {http://arxiv.org/abs/0901.3753} {\path{arXiv:0901.3753}},
  \href {http://dx.doi.org/10.1103/PhysRevLett.103.131601}
  {\path{doi:10.1103/PhysRevLett.103.131601}}.

\bibitem{Arutyunov:2009zu}
G.~Arutyunov, S.~Frolov, {String hypothesis for the $ AdS_5 \times S^5 $
  mirror}, JHEP 03 (2009) 152.
\newblock \href {http://arxiv.org/abs/0901.1417} {\path{arXiv:0901.1417}},
  \href {http://dx.doi.org/10.1088/1126-6708/2009/03/152}
  {\path{doi:10.1088/1126-6708/2009/03/152}}.

\bibitem{Bombardelli:2009ns}
D.~Bombardelli, D.~Fioravanti, R.~Tateo, {Thermodynamic Bethe Ansatz for planar
  AdS/CFT: a proposal}, J. Phys. A42 (2009) 375401.
\newblock \href {http://arxiv.org/abs/0902.3930} {\path{arXiv:0902.3930}},
  \href {http://dx.doi.org/10.1088/1751-8113/42/37/375401}
  {\path{doi:10.1088/1751-8113/42/37/375401}}.

\bibitem{Arutyunov:2009ur}
G.~Arutyunov, S.~Frolov, {Thermodynamic Bethe Ansatz for the $AdS_5 x S^5$
  Mirror Model}, JHEP 05 (2009) 068.
\newblock \href {http://arxiv.org/abs/0903.0141} {\path{arXiv:0903.0141}},
  \href {http://dx.doi.org/10.1088/1126-6708/2009/05/068}
  {\path{doi:10.1088/1126-6708/2009/05/068}}.

\bibitem{Gromov:2009bc}
N.~Gromov, V.~Kazakov, A.~Kozak, P.~Vieira, {Exact Spectrum of Anomalous
  Dimensions of Planar N = 4 Supersymmetric Yang-Mills Theory: TBA and excited
  states}, Lett. Math. Phys. 91 (2010) 265--287.
\newblock \href {http://arxiv.org/abs/0902.4458} {\path{arXiv:0902.4458}},
  \href {http://dx.doi.org/10.1007/s11005-010-0374-8}
  {\path{doi:10.1007/s11005-010-0374-8}}.

\bibitem{Arutyunov:2009ax}
G.~Arutyunov, S.~Frolov, R.~Suzuki, {Exploring the mirror TBA}, JHEP 05 (2010)
  031.
\newblock \href {http://arxiv.org/abs/0911.2224} {\path{arXiv:0911.2224}},
  \href {http://dx.doi.org/10.1007/JHEP05(2010)031}
  {\path{doi:10.1007/JHEP05(2010)031}}.

\bibitem{Cavaglia:2010nm}
A.~Cavaglia, D.~Fioravanti, R.~Tateo, {Extended Y-system for the $AdS_5/CFT_4$
  correspondence, }\href {http://arxiv.org/abs/1005.3016}
  {\path{arXiv:1005.3016}}.

\bibitem{Arutyunov:2010gb}
G.~Arutyunov, S.~Frolov, R.~Suzuki, {Five-loop Konishi from the Mirror TBA},
  JHEP 04 (2010) 069.
\newblock \href {http://arxiv.org/abs/1002.1711} {\path{arXiv:1002.1711}},
  \href {http://dx.doi.org/10.1007/JHEP04(2010)069}
  {\path{doi:10.1007/JHEP04(2010)069}}.

\bibitem{Balog:2010xa}
J.~Balog, A.~Hegedus, {5-loop Konishi from linearized TBA and the XXX magnet},
  JHEP 06 (2010) 080.
\newblock \href {http://arxiv.org/abs/1002.4142} {\path{arXiv:1002.4142}},
  \href {http://dx.doi.org/10.1007/JHEP06(2010)080}
  {\path{doi:10.1007/JHEP06(2010)080}}.

\bibitem{Balog:2010vf}
J.~Balog, A.~Hegedus, {The Bajnok-Janik formula and wrapping corrections,
  }\href {http://arxiv.org/abs/1003.4303} {\path{arXiv:1003.4303}}.

\bibitem{Arutyunov:2010gu}
G.~Arutyunov, M.~de~Leeuw, S.~J. van Tongeren, {Twisting the Mirror TBA, }\href
  {http://arxiv.org/abs/1009.4118} {\path{arXiv:1009.4118}}.

\bibitem{Mann:2006rh}
N.~Mann, S.~E. Vazquez, {Classical open string integrability}, JHEP 04 (2007)
  065.
\newblock \href {http://arxiv.org/abs/hep-th/0612038}
  {\path{arXiv:hep-th/0612038}}.

\bibitem{Berenstein:2005vf}
D.~Berenstein, S.~E. Vazquez, {Integrable open spin chains from giant
  gravitons}, JHEP 06 (2005) 059.
\newblock \href {http://arxiv.org/abs/hep-th/0501078}
  {\path{arXiv:hep-th/0501078}}.

\bibitem{Hofman:2007xp}
D.~M. Hofman, J.~M. Maldacena, {Reflecting magnons}, JHEP 11 (2007) 063.
\newblock \href {http://arxiv.org/abs/0708.2272} {\path{arXiv:0708.2272}},
  \href {http://dx.doi.org/10.1088/1126-6708/2007/11/063}
  {\path{doi:10.1088/1126-6708/2007/11/063}}.

\bibitem{Galleas:2009ye}
W.~Galleas, {The Bethe Ansatz Equations for Reflecting Magnons}, Nucl. Phys.
  B820 (2009) 664--681.
\newblock \href {http://arxiv.org/abs/0902.1681} {\path{arXiv:0902.1681}},
  \href {http://dx.doi.org/10.1016/j.nuclphysb.2009.04.024}
  {\path{doi:10.1016/j.nuclphysb.2009.04.024}}.

\bibitem{Nepomechie:2009zi}
R.~I. Nepomechie, {Bethe ansatz equations for open spin chains from giant
  gravitons}, JHEP 05 (2009) 100.
\newblock \href {http://arxiv.org/abs/0903.1646} {\path{arXiv:0903.1646}},
  \href {http://dx.doi.org/10.1088/1126-6708/2009/05/100}
  {\path{doi:10.1088/1126-6708/2009/05/100}}.

\bibitem{Correa:2009dm}
D.~H. Correa, C.~A.~S. Young, {Asymptotic Bethe equations for open boundaries
  in planar AdS/CFT}, J. Phys. A43 (2010) 145401.
\newblock \href {http://arxiv.org/abs/0912.0627} {\path{arXiv:0912.0627}},
  \href {http://dx.doi.org/10.1088/1751-8113/43/14/145401}
  {\path{doi:10.1088/1751-8113/43/14/145401}}.

\bibitem{Palla:2008zc}
L.~Palla, {Issues on magnon reflection}, Nucl. Phys. B808 (2009) 205--223.
\newblock \href {http://arxiv.org/abs/0807.3646} {\path{arXiv:0807.3646}},
  \href {http://dx.doi.org/10.1016/j.nuclphysb.2008.09.021}
  {\path{doi:10.1016/j.nuclphysb.2008.09.021}}.

\bibitem{Ghoshal:1993tm}
S.~Ghoshal, A.~B. Zamolodchikov, {Boundary S matrix and boundary state in
  two-dimensional integrable quantum field theory}, Int. J. Mod. Phys. A9
  (1994) 3841--3886.
\newblock \href {http://arxiv.org/abs/hep-th/9306002}
  {\path{arXiv:hep-th/9306002}}, \href
  {http://dx.doi.org/10.1142/S0217751X94001552}
  {\path{doi:10.1142/S0217751X94001552}}.

\bibitem{Correa:2009mz}
D.~H. Correa, C.~A.~S. Young, {Finite size corrections for open strings/open
  chains in planar AdS/CFT}, JHEP 08 (2009) 097.
\newblock \href {http://arxiv.org/abs/0905.1700} {\path{arXiv:0905.1700}},
  \href {http://dx.doi.org/10.1088/1126-6708/2009/08/097}
  {\path{doi:10.1088/1126-6708/2009/08/097}}.

\bibitem{Bajnok:2004tq}
Z.~Bajnok, L.~Palla, G.~Takacs, {Finite size effects in quantum field theories
  with boundary from scattering data}, Nucl. Phys. B716 (2005) 519--542.
\newblock \href {http://arxiv.org/abs/hep-th/0412192}
  {\path{arXiv:hep-th/0412192}}, \href
  {http://dx.doi.org/10.1016/j.nuclphysb.2005.03.021}
  {\path{doi:10.1016/j.nuclphysb.2005.03.021}}.

\bibitem{Bajnok:2006dn}
Z.~Bajnok, L.~Palla, G.~Takacs, {Boundary one-point function, Casimir energy
  and boundary state formalism in D+1 dimensional QFT}, Nucl. Phys. B772 (2007)
  290--322.
\newblock \href {http://arxiv.org/abs/hep-th/0611176}
  {\path{arXiv:hep-th/0611176}}, \href
  {http://dx.doi.org/10.1016/j.nuclphysb.2007.02.023}
  {\path{doi:10.1016/j.nuclphysb.2007.02.023}}.

\bibitem{Balasubramanian:2002sa}
V.~Balasubramanian, M.-x. Huang, T.~S. Levi, A.~Naqvi, {Open strings from N = 4
  super Yang-Mills}, JHEP 08 (2002) 037.
\newblock \href {http://arxiv.org/abs/hep-th/0204196}
  {\path{arXiv:hep-th/0204196}}.

\bibitem{Chen:2007ec}
H.-Y. Chen, D.~H. Correa, {Comments on the Boundary Scattering Phase}, JHEP 02
  (2008) 028.
\newblock \href {http://arxiv.org/abs/0712.1361} {\path{arXiv:0712.1361}},
  \href {http://dx.doi.org/10.1088/1126-6708/2008/02/028}
  {\path{doi:10.1088/1126-6708/2008/02/028}}.

\bibitem{Ahn:2008df}
C.~Ahn, R.~I. Nepomechie, {The Zamolodchikov-Faddeev algebra for open strings
  attached to giant gravitons}, JHEP 05 (2008) 059.
\newblock \href {http://arxiv.org/abs/0804.4036} {\path{arXiv:0804.4036}},
  \href {http://dx.doi.org/10.1088/1126-6708/2008/05/059}
  {\path{doi:10.1088/1126-6708/2008/05/059}}.

\bibitem{Arutyunov:2009kf}
G.~Arutyunov, S.~Frolov, {The Dressing Factor and Crossing Equations}, J. Phys.
  A42 (2009) 425401.
\newblock \href {http://arxiv.org/abs/0904.4575} {\path{arXiv:0904.4575}},
  \href {http://dx.doi.org/10.1088/1751-8113/42/42/425401}
  {\path{doi:10.1088/1751-8113/42/42/425401}}.

\bibitem{Beisert:2006ez}
N.~Beisert, B.~Eden, M.~Staudacher, {Transcendentality and crossing}, J. Stat.
  Mech. 0701 (2007) P021.
\newblock \href {http://arxiv.org/abs/hep-th/0610251}
  {\path{arXiv:hep-th/0610251}}.

\bibitem{Luscher:1985dn}
M.~Luscher, {Volume Dependence of the Energy Spectrum in Massive Quantum Field
  Theories. 1. Stable Particle States}, Commun. Math. Phys. 104 (1986) 177.
\newblock \href {http://dx.doi.org/10.1007/BF01211589}
  {\path{doi:10.1007/BF01211589}}.

\bibitem{Gromov:2008gj}
N.~Gromov, V.~Kazakov, P.~Vieira, {Finite Volume Spectrum of 2D Field Theories
  from Hirota Dynamics}, JHEP 12 (2009) 060.
\newblock \href {http://arxiv.org/abs/0812.5091} {\path{arXiv:0812.5091}},
  \href {http://dx.doi.org/10.1088/1126-6708/2009/12/060}
  {\path{doi:10.1088/1126-6708/2009/12/060}}.

\bibitem{Balog:2009ze}
J.~Balog, A.~Hegedus, {The finite size spectrum of the 2-dimensional O(3)
  nonlinear sigma-model, }\href {http://arxiv.org/abs/0907.1759}
  {\path{arXiv:0907.1759}}.

\bibitem{Arutyunov:2007tc}
G.~Arutyunov, S.~Frolov, {On String S-matrix, Bound States and TBA}, JHEP 12
  (2007) 024.
\newblock \href {http://arxiv.org/abs/0710.1568} {\path{arXiv:0710.1568}},
  \href {http://dx.doi.org/10.1088/1126-6708/2007/12/024}
  {\path{doi:10.1088/1126-6708/2007/12/024}}.

\bibitem{Sklyanin:1988yz}
E.~K. Sklyanin, {Boundary Conditions for Integrable Quantum Systems, }, J.
  Phys. A21 (1988) 2375.
\newblock \href {http://dx.doi.org/10.1088/0305-4470/21/10/015}
  {\path{doi:10.1088/0305-4470/21/10/015}}.

\bibitem{Ahn:2000jd}
C.-r. Ahn, R.~I. Nepomechie, {Exact solution of the supersymmetric sinh-Gordon
  model with boundary}, Nucl. Phys. B586 (2000) 611--640.
\newblock \href {http://arxiv.org/abs/hep-th/0005170}
  {\path{arXiv:hep-th/0005170}}, \href
  {http://dx.doi.org/10.1016/S0550-3213(00)00440-5}
  {\path{doi:10.1016/S0550-3213(00)00440-5}}.

\bibitem{Pearce:2000dv}
P.~A. Pearce, L.~Chim, C.-r. Ahn, {Excited TBA equations. I: Massive
  tricritical Ising model}, Nucl. Phys. B601 (2001) 539--568.
\newblock \href {http://arxiv.org/abs/hep-th/0012223}
  {\path{arXiv:hep-th/0012223}}, \href
  {http://dx.doi.org/10.1016/S0550-3213(01)00081-5}
  {\path{doi:10.1016/S0550-3213(01)00081-5}}.

\bibitem{Dorey:1997yg}
P.~Dorey, A.~Pocklington, R.~Tateo, G.~Watts, {TBA and TCSA with boundaries and
  excited states}, Nucl. Phys. B525 (1998) 641--663.
\newblock \href {http://arxiv.org/abs/hep-th/9712197}
  {\path{arXiv:hep-th/9712197}}, \href
  {http://dx.doi.org/10.1016/S0550-3213(98)00339-3}
  {\path{doi:10.1016/S0550-3213(98)00339-3}}.

\bibitem{Bajnok:2007ep}
Z.~Bajnok, C.~Rim, A.~Zamolodchikov, {Sinh-Gordon Boundary TBA and Boundary
  Liouville Reflection Amplitude}, Nucl. Phys. B796 (2008) 622--650.
\newblock \href {http://arxiv.org/abs/0710.4789} {\path{arXiv:0710.4789}},
  \href {http://dx.doi.org/10.1016/j.nuclphysb.2007.12.023}
  {\path{doi:10.1016/j.nuclphysb.2007.12.023}}.

\bibitem{Ahn:2003st}
C.~Ahn, M.~Bellacosa, F.~Ravanini, {Excited states NLIE for sine-Gordon model
  in a strip with Dirichlet boundary conditions}, Phys. Lett. B595 (2004)
  537--546.
\newblock \href {http://arxiv.org/abs/hep-th/0312176}
  {\path{arXiv:hep-th/0312176}}, \href
  {http://dx.doi.org/10.1016/j.physletb.2004.04.007}
  {\path{doi:10.1016/j.physletb.2004.04.007}}.

\bibitem{Janik:2006dc}
R.~A. Janik, {The AdS(5) x S**5 superstring worldsheet S-matrix and crossing
  symmetry}, Phys. Rev. D73 (2006) 086006.
\newblock \href {http://arxiv.org/abs/hep-th/0603038}
  {\path{arXiv:hep-th/0603038}}, \href
  {http://dx.doi.org/10.1103/PhysRevD.73.086006}
  {\path{doi:10.1103/PhysRevD.73.086006}}.

\bibitem{Arutyunov:2008zt}
G.~Arutyunov, S.~Frolov, {The S-matrix of String Bound States}, Nucl. Phys.
  B804 (2008) 90--143.
\newblock \href {http://arxiv.org/abs/0803.4323} {\path{arXiv:0803.4323}},
  \href {http://dx.doi.org/10.1016/j.nuclphysb.2008.06.005}
  {\path{doi:10.1016/j.nuclphysb.2008.06.005}}.

\bibitem{Ahn:2010xa}
C.~Ahn, R.~I. Nepomechie, {Yangian symmetry and bound states in AdS/CFT
  boundary scattering}, JHEP 05 (2010) 016.
\newblock \href {http://arxiv.org/abs/1003.3361} {\path{arXiv:1003.3361}},
  \href {http://dx.doi.org/10.1007/JHEP05(2010)016}
  {\path{doi:10.1007/JHEP05(2010)016}}.

\bibitem{MacKay:2010ey}
N.~MacKay, V.~Regelskis, {Yangian symmetry of the Y=0 maximal giant graviton,
  }\href {http://arxiv.org/abs/1010.3761} {\path{arXiv:1010.3761}}.

\bibitem{Murgan:2008fs}
R.~Murgan, R.~I. Nepomechie, {Open-chain transfer matrices for AdS/CFT}, JHEP
  09 (2008) 085.
\newblock \href {http://arxiv.org/abs/0808.2629} {\path{arXiv:0808.2629}},
  \href {http://dx.doi.org/10.1088/1126-6708/2008/09/085}
  {\path{doi:10.1088/1126-6708/2008/09/085}}.

\bibitem{Ahn:2007bq}
C.~Ahn, D.~Bak, S.-J. Rey, {Reflecting Magnon Bound States}, JHEP 04 (2008)
  050.
\newblock \href {http://arxiv.org/abs/0712.4144} {\path{arXiv:0712.4144}},
  \href {http://dx.doi.org/10.1088/1126-6708/2008/04/050}
  {\path{doi:10.1088/1126-6708/2008/04/050}}.

\bibitem{Sieg:2005kd}
C.~Sieg, A.~Torrielli, {Wrapping interactions and the genus expansion of the 2-
  point function of composite operators}, Nucl. Phys. B723 (2005) 3--32.
\newblock \href {http://arxiv.org/abs/hep-th/0505071}
  {\path{arXiv:hep-th/0505071}}, \href
  {http://dx.doi.org/10.1016/j.nuclphysb.2005.06.011}
  {\path{doi:10.1016/j.nuclphysb.2005.06.011}}.

\bibitem{Ahn:2010yv}
C.~Ahn, Z.~Bajnok, D.~Bombardelli, R.~I. Nepomechie, {Finite-size effect for
  four-loop Konishi of the beta- deformed N=4 SYM}, Phys. Lett. B693 (2010)
  380--385.
\newblock \href {http://arxiv.org/abs/1006.2209} {\path{arXiv:1006.2209}},
  \href {http://dx.doi.org/10.1016/j.physletb.2010.08.056}
  {\path{doi:10.1016/j.physletb.2010.08.056}}.

\bibitem{Beisert:2005if}
N.~Beisert, R.~Roiban, {Beauty and the twist: The Bethe ansatz for twisted N =
  4 SYM}, JHEP 08 (2005) 039.
\newblock \href {http://arxiv.org/abs/hep-th/0505187}
  {\path{arXiv:hep-th/0505187}}.

\bibitem{Gromov:2010dy}
N.~Gromov, F.~Levkovich-Maslyuk, {Y-system and beta-deformed N=4
  Super-Yang-Mills, }\href {http://arxiv.org/abs/1006.5438}
  {\path{arXiv:1006.5438}}.

\bibitem{Ahn:2010ws}
C.~Ahn, Z.~Bajnok, D.~Bombardelli, R.~I. Nepomechie, {Twisted Bethe equations
  from a twisted S-matrix, }\href {http://arxiv.org/abs/1010.3229}
  {\path{arXiv:1010.3229}}.

\end{thebibliography}
\end{document}